\documentclass[longabstract,manuscript]{aastex}

\usepackage[totalwidth=520pt, totalheight=680pt, left=15mm, letterpaper]{geometry}

\usepackage[OT4]{fontenc}

\DeclareMathAlphabet{\mathitbf}{OML}{cmm}{b}{it}
\newcommand{\de}{\mathrm{d}} 
\newcommand{\ind}{\indent}

\newcommand{\etal}{{et~al.\ }}
\newcommand{\Mpc}{\;\mathrm{Mpc}}
\newcommand{\Mpch}{\;\mathrm{Mpc} / h}
\newcommand{\hMpc}{\;h / \mathrm{Mpc}}
\newcommand{\kms}{\;\mathrm{km} / \mathrm{s}}
\newcommand{\bmd}{\mathitbf{d}}
\newcommand{\bmg}{\mathitbf{g}}
\newcommand{\bmk}{\mathitbf{k}}
\newcommand{\bmr}{\mathitbf{r}}
\newcommand{\bmu}{\mathitbf{u}}
\newcommand{\bmv}{\mathitbf{v}}

\newcommand{\mrc}{\mathrm{c}}
\newcommand{\mrm}{\mathrm{m}}
\newcommand{\mrg}{\mathrm{g}}

\newcommand{\lan}{\langle}
\newcommand{\ran}{\rangle}
\newcommand{\Omm}{\Omega_\mathrm{m}} 
\newcommand{\Omb}{\Omega_\mathrm{b}}
\newcommand{\dgr}{^{\circ}}

\shorttitle{Convergence of 2MASS dipole}
\shortauthors{Bilicki \etal}

\begin{document}

\title{Is the 2MASS clustering dipole convergent?}
\author{Maciej Bilicki}
\affil{Nicolaus Copernicus Astronomical Center, Polish Academy of Sciences, \\ul.\ Bartycka 18, 00-716, Warsaw, Poland}
\email{bilicki(at)camk.edu.pl}

\author{Micha{\l} Chodorowski}
\affil{Nicolaus Copernicus Astronomical Center, Polish Academy of Sciences, \\ul.\ Bartycka 18, 00-716, Warsaw, Poland}

\author{Thomas Jarrett}
\affil{Spitzer Science Center, California Institute of Technology, \\Pasadena, CA 91125, USA}

\and

\author{Gary A. Mamon}
\affil{Institut d'Astrophysique de Paris (UMR 7095: CNRS \& UPMC), \\98 bis Bd Arago, F-75014 Paris, France}
\begin{abstract}
There is a long-standing controversy about the convergence of the dipole moment of the galaxy angular distribution (the so-called \emph{clustering dipole}). Is the dipole convergent at all, and if so, what is the scale of the convergence? We study the growth of the clustering dipole of galaxies as a function of the limiting flux of the sample from the Two Micron All Sky Survey (2MASS). Contrary to some earlier claims, we find that the dipole does not converge before the completeness limit of the 2MASS Extended Source Catalog, i.e.\ up to $13.5$ mag in the near-infrared $K_s$ band (equivalent to an effective distance of $300\Mpch$). We compare the observed growth of the dipole with the theoretically expected, conditional one (i.e., given the velocity of the Local Group relative to the CMB), for the $\Lambda$CDM power spectrum and cosmological parameters constrained by \emph{WMAP}. The observed growth turns out to be within 1$\sigma$ confidence level of its theoretical counterpart once the proper observational window of the 2MASS flux-limited catalog is included. For a contrast, if the adopted window is a top-hat, then the predicted dipole grows significantly faster and converges (within the errors) to its final value for a distance of about $300\Mpch$. By comparing the observational windows, we show that for a given flux limit and a corresponding distance limit, the 2MASS flux-weighted window passes less large-scale signal than the top-hat one. We conclude that the growth of the 2MASS dipole for effective distances greater than $200\Mpch$ is only apparent. On the other hand, for a distance of $80\Mpch$ (mean depth of the 2MASS Redshift Survey) and the $\Lambda$CDM power spectrum, the true dipole is expected to reach only $\sim80\%$ of its final value. Eventually, since for the window function of 2MASS the predicted growth is consistent with the observed one, we can compare the two to evaluate $\beta\equiv \Omm^{0.55}\slash b$. The result is $\beta = 0.38\pm0.04$, which leads to an estimate of the density parameter $\Omm=0.20\pm0.08$.
\end{abstract}

%
%
%
\section{Introduction}\label{Sec:Introduction} 
The currently accepted paradigm of large-scale structure formation in the Universe is the gravitational instability. Within this framework, inhomogeneities in matter distribution induce gravitational accelerations, which result in galaxies having \emph{peculiar velocities} that add to the Hubble flow. These velocities in turn enhance the growth of the inhomogeneities, causing strong coupling between cosmic velocity field and large-scale matter distribution. In perturbation theory of Friedman-Lema\^{\i}tre models, in the linear regime, peculiar velocities and accelerations are aligned and proportional to each other at every point (e.g.\ \citealt{Pe80}). More importantly, the proportionality coefficient of this relation is a simple function of the cosmological parameter of non-relativistic matter density, $\Omm$, and practically does not depend on the cosmological constant (nor other forms of dark energy). Additionally, as the relation holds also for the motion of the whole \emph{Local Group of galaxies} (LG) through the Universe, measurements of its peculiar velocity and acceleration may be used as a tool to dynamically constrain the $\Omm$ parameter.

The peculiar velocity of the LG is known from the observed dipole anisotropy of the cosmic microwave background (CMB, \citealt{Hinsh}), interpreted as a kinematic effect, and reduced to the barycenter of the LG \citep{CvdB}. It equals to $v_\mathrm{CMB}=622\pm35\kms$ and points in the direction $(l,b)=(272\dgr\pm3\dgr\!,\,28\dgr\pm5\dgr)$ in Galactic coordinates. This kinematic interpretation is strongly supported by the observed alignment of the CMB dipole with the direction of the peculiar acceleration of the LG, although the latter is much more difficult to estimate. It requires knowledge of mass distribution in our cosmic neighborhood, and its determination had not been possible until deep all-sky galaxy catalogs became available. For that reason, the first attempts to measure the acceleration of the LG were made not earlier than 30 years ago \citep{YST, DH82}.

Using an all-sky catalog, such measurement can be made under the assumption that visible (luminous) matter is a good tracer of the underlying density field. The general procedure is to estimate the so-called \emph{clustering dipole} of a galaxy survey and infer the acceleration of the LG. However, such inference requires several conditions to be met. First, the survey should cover the whole sky; second, the observational proxy of the gravitational force (most often the flux of the galaxy in the photometric band of the survey) should have known properties; and last but not least, the survey should be deep enough for the dipole to converge to the final value that we want to find. As usually one or more of these assumptions do not hold, the clustering dipole is a biased estimator of the acceleration and the estimation of the latter from the former may be done only if the mentioned effects are properly accounted for.

In the past, many different datasets have been applied to calculate the clustering dipole. Generally speaking, there is no consistency on the \emph{amplitude}, the \emph{scale of convergence} of the dipole, and even on the \textit{convergence itself}. The pioneering works used the revised Shapley-Ames catalog \citep{YST} and the CfA catalog \citep{DH82}. A great advancement came with the launch of the far-infrared \emph{IRAS} satellite and catalogs obtained thanks to this mission. The LG dipole from \emph{IRAS} was studied first from two-dimensional data only \citep{YWRR, MD86, HLM, ViSt}, then with redshifts included \citep{Strauss, Schmoldt, RR2000, DMSa, BP06} and with optical data added \citep{LRRLB, KaLa}. Samples with optical data only were also used \citep{Lah87, Hudson}, as well as galaxy clusters \citep{Tini, KE06}. Among the most recent analyses one finds those directly related to the study presented here, which used the data from the Two Micron All Sky Survey (2MASS, \citealt{Skr}; see also hereunder). \cite{Maller03} used the 2MASS Extended Source Catalog (XSC, \citealt{Jar00}), concluded convergence of the clustering dipole from flux data only and used it to calculate the average mass-to-light ratio in the near-infrared $K_s$ band and to estimate the linear biasing parameter $b_L$. \cite{Erdogdu} studied the acceleration of the LG from the 2MASS Redshift Survey (2MRS, an all-sky subset of the 2MASS catalog with measured redshifts, \citealt{Huchra}) and estimated a combination of the mass density $\Omm$ and luminosity bias parameters, the so-called $\beta$ parameter. In a recent work, \cite{Lav10} used an orbit-reconstruction algorithm to generate the peculiar velocity field for the 2MRS, extended it to larger radii, and observed no convergence up to at least $120\Mpch$. 

Since both the gravitational force and the flux of a galaxy are inversely proportional to the distance squared, the clustering dipole can be calculated by assuming some average value of the mass-to-light ratio and summing only flux `vectors' of galaxies (with positions on the sky used as angular coordinates). Such a dipole does not use distance nor redshift information of particular objects, which hinders the estimation of the depth up to which the dipole is measured and can bias the estimation of LG acceleration from it. This drawback is partially removed if a galaxy \emph{redshift} survey (with redshifts of galaxies used as proxies of their distances) is used for the analysis; however, up to date the densest and deepest \emph{all-sky} redshift surveys do not reach farther than to $\sim100\Mpch$ and contain no more than several dozen thousand sources. Owing to some recent claims of large-scale flows in the local Universe \citep{WFH09,KABK}, it is important to estimate the dipole from as big catalogs as possible, even if they do not contain redshift measurements. We thus have decided to trade the advantages of redshift measurements for a huge number of galaxies and a much greater depth of a photometric-only catalog. The angular dipole thus obtained is additionally free from any so-called \emph{redshift distortions}, in particular from the \emph{rocket effect} \citep{Kaiser}.

In this Paper, we focus mainly on the issue of the convergence of the dipole, by analyzing the growth of the dipole with increased depth of the sample. For that purpose we use the data from the 2MASS XSC, similarly to what has been presented in \cite{Maller03}. One of the goals of this Paper is to expand and refine the latter work. We do it by modifying the criteria of removing Galactic objects from the sample and masking and filling the Galactic plane and bulge (the Zone of Avoidance, ZoA), but keeping in the ZoA the brightest and closest galaxies of big influence for the local motion. We then analyze the growth of the dipole and check how its direction changes on the sky. Next, we compare the observed growth of the dipole with theoretical expectations, using the condition of known peculiar velocity of the Local Group, in a similar manner as was done in \cite{Lav10}. From this comparison we evaluate the combination of the cosmological matter density parameter and the linear bias (the `$\beta$' parameter) and, by using externally provided value of bias, constrain the parameter $\Omm$. 

The paper is organized as follows. In Section \ref{Sec:Dipole}, we present the theoretical framework for the calculation of the clustering dipole. Section \ref{Sec:Data} contains details on how we handled the 2MASS data. Next, Section \ref{Sec:Growth} focuses on the growth of the clustering dipole. Section \ref{Subsec:Angle} focuses on the position of the 2MASS dipole on the sky. In subsequent Section \ref{Sec:Discussion} we discuss the results: first, in Subsection \ref{Subsec:Theory} we present the theoretical basis to analyze the growth for known LG velocity; Subsection \ref{Subsec:Windows} focuses on the observational window of our measurement; finally in Subsection \ref{Subsec:Compare} we compare our results with theoretical expectations and use this comparison to estimate the $\beta$ parameter from the growth of the dipole alone. We shortly summarize and conclude in Section \ref{Sec:SummConcl}, where we also provide the obtained value of the $\Omm$ parameter.

\section{The clustering dipole}\label{Sec:Dipole} 
The gravitational instability scenario of large-scale structure formation relates peculiar velocities of galaxies with their peculiar gravitational accelerations. In linear theory, this relation has a particularly simple form \cite{Pe80}:
\begin{equation}\label{eq:v.and.g}
\bmv=\frac{H_0 f(\Omega_\mrm)}{4 \pi G \rho_b}\;\bmg=\frac{2 f(\Omega_\mrm)}{3 H_0 \Omega_\mrm}\;\bmg\;.  
\end{equation}
Here, $H_0=100\,h\,\kms / \mathrm{Mpc}$ is the Hubble constant, $\Omega_\mrm$ is the current value of the co\-s\-mo\-lo\-gi\-cal density parameter of non-relativistic matter, $f(\Omega_\mrm)\equiv\left(\de\ln D\slash\de \ln z\right)\vert_{z=0}$ (with $D$ being the growth factor) and $\rho_b$ is the mean matter density of the background. Within $\Lambda$CDM models with a cosmological constant, the growth parameter $f$ is very well fitted by $f(\Omega_\mrm)\simeq\Omega_\mrm^{0.55}$ \citep{Linder} and is virtually independent of $\Lambda$, cf.\ \cite{LLPR}. The acceleration vector at a position `$\bmr$' is given by the integral: 
\begin{equation}\label{eq:g.theor}
\bmg(\bmr)=G \rho_b \int{\delta_\mrm(\bmr ')\,\frac{\bmr '-\bmr }{|\bmr '-\bmr |^3}\,\de^3 \bmr '}\;,   
\end{equation}
where $\delta_\mrm(\bmr)=\left[\rho_\mrm(\bmr)-\rho_b\right]\slash \rho_b$ is the density contrast of non-relativistic matter at the point $\bmr$. However, as what we observe are galaxies, we have to assume some relation between their density field and that of matter. This is usually done via the \emph{linear biasing} paradigm: $\delta_\mrg=b\, \delta_\mrm$. This biasing scheme, valid in linear theory that we use in our whole analysis, neglects the stochasticity, as well as possible scale- and galaxy-type dependence in the relation between the two density fields. For more details on possible non-linear biasing, see for example the review by \cite{LaSu}.

Including the biasing relation into Eq.\ (\ref{eq:g.theor}) and using the fact that for a spherical survey $\int\frac{\bmr '-\bmr}{|\bmr '-\bmr |^{3}}\,\de^3 \bmr '=0$, we get the following expression for the peculiar acceleration:
\begin{equation}\label{eq:g.galaxies}
\bmg(\bmr)=\frac{G}{b}\int{\rho_\mrg(\bmr ')\,\frac{\bmr '-\bmr }{|\bmr '-\bmr |^3}\,\de^3 \bmr '}\;.
\end{equation}
The biasing parameter $b$ is usually combined with the factor $f(\Omega_\mrm)$ into the parameter $\beta\equiv f(\Omega_\mrm)\slash b$. Comparing Eqs.\ (\ref{eq:v.and.g}) and (\ref{eq:g.galaxies}), we get the proportionality valid in linear theory:
\begin{equation}
\bmv \propto  \beta\, \bmg\;.
\end{equation}
This relation could in principle be applied directly to the motion of the Local Group of galaxies through the Universe. Consequently, comparison of the peculiar velocity and acceleration of the LG serves as a tool to estimate the $\beta$ parameter. Independent knowledge of biasing allows to estimate cosmological density $\Omm$.

In reality, however, we do not observe continuous galaxy density field, but rather discrete objects, even if in a very large number. In the following derivation, which can be found e.g.\ in \cite{ViSt}, we model galaxies as point sources: $\rho_\mrg(\bmr) = \sum_i M_i\, \delta_D(\bmr - \bmr_i)$, where $\delta_D$ is Dirac's delta; $M_i$ and $\bmr_i$ are respectively the mass and the position of the $i$-th galaxy. Putting the coordinate system at $\bmr=0$ (barycenter of the LG) we obtain the acceleration of the LG as a sum of the force contributions from all sources in the Universe:
\begin{equation}
\bmg=\frac{G}{b} \sum_i M_i \frac{\hat{\bmr}_i}{r^2_i}.
\end{equation}
This Newtonian formula\footnote{The Newtonian limit can be applied as our whole analysis concerns distances well below the Hubble radius $r_H\equiv c\slash H_0=3\, \mathrm{Gpc} / h$.} is still not useful for calculations based on observational data, as masses of individual galaxies are usually known very poorly, if ever. However, if the $i$-th galaxy has an intrinsic luminosity $L_i$, we can write
\begin{equation}
\bmg=\frac{4\pi G}{b} \sum_i \frac{M_i}{L_i}\frac{L_i}{4\pi r_i^2}\hat{\bmr}_i=\frac{4\pi G}{b} \sum_i \frac{M_i}{L_i}S_i\hat{\bmr}_i\;,
\end{equation}
where $S_i=L_i\slash 4\pi r_i^2$ is the flux received from the $i$-th object. This relation means that if we know the behavior of the mass-to-light ratio in the band(s) of the survey, we can estimate the acceleration of the LG from a \emph{two-dimensional} catalog, i.e.\ one containing astro- and photometric data only (positions and fluxes). Furthermore, if the mean mass-to-light ratio is a universal constant, \mbox{$\Upsilon=\lan M\slash L\ran$}, we finally get\footnote{Note that as we are interested in the motion of the Local Group as an entire system, the galaxies of the LG should \emph{not} be included in the summation.}
\begin{equation}
\bmg=\frac{4\,\pi\,G\,\Upsilon}{b} \sum_i S_i\hat{\bmr}_i\;.
\end{equation}
In some applications, including the present one, it is more convenient to work in terms of matter and luminosity densities. This is especially the case when the \emph{luminosity density}, $j$, is known for a given band, rather than the mass-to-light ratio. We have
\begin{equation}
\Upsilon=\left\lan{\frac{M}{L}}\right\ran=\frac{\rho_\mrm}{j}=\frac{3\, \Omega_\mrm\,H_0^2}{8\,\pi\, G\, j}\;
\end{equation}
which gives
\begin{equation}\label{eq:flux dipole}
\bmg=\frac{3\, \Omega_\mrm \, H_0^2}{2\,b\,j}\sum_i S_i\hat{\bmr}_i\;.
\end{equation}
The luminosity density $j$ for a particular band of the survey can be calculated for example from the luminosity function $\Phi(L)$ of galaxies in this band (e.g.\ \citealt{Pe93})
\begin{equation}\label{eq:lum.dens} 
j=\int\limits_0^{\infty}{L\, \Phi(L)\, \de L }\;.
\end{equation}
Note that using the Relation (\ref{eq:flux dipole}) in Eq.\ (\ref{eq:v.and.g}), we get the linear-theory velocity measured from the flux dipole as
\begin{equation}\label{eq:v_s}
\bmv=\beta\, \frac{H_0}{j}\sum_i S_i\hat{\bmr}_i=\beta\, \tilde{\bmg}\;,
\end{equation}
where $\tilde{\bmg}$ denotes the \emph{scaled} acceleration of the LG, in units of velocity. The term $\sum_i S_i\hat{\bmr}_i$ is the flux dipole moment of \emph{all} sources down to the zero flux over the whole sky. The universal luminosity density $j$, measured from a fair sample of galaxies in the given band with known apparent luminosities and redshifts, is proportional to $H_0$, which means that the overall result does not depend on the Hubble constant.

Realistic galaxy catalogs will never reach down to zero flux, irrespectively of the used wavelength. On the contrary, surveys are usually \emph{flux-limited}, which means that the number of observed sources, $N$, is finite. For that reason, in the following we will denote the flux dipole of a finite sample as $\bmd$:
\begin{equation}\label{eq:lim flux dipole}
\bmd=\frac{3\, \Omega_\mrm \, H_0^2}{2\,b\,j}\sum_i^N S_i\hat{\bmr}_i\;.
\end{equation}
Note that the clustering dipole calculated for a finite, flux-limited sample may be a biased estimator of the peculiar acceleration of the Local Group, Eq.\ (\ref{eq:g.theor}). This can be overcome by extrapolating the measured dipole to zero flux \citep{ViSt}. The situation gets easier in the case of galaxy \emph{redshift} surveys. We can then use the redshifts as distance estimators and weight galaxies with the inverse of the selection function of the survey (e.g.\ \citealt{YSDH}), in order to mimic an ideal, volume-limited catalog. However, despite an outstanding advancement in surveying the cosmos in recent years, the deepest and densest all-sky redshift survey, 2MRS, contains only $\sim$ 43,000 galaxies and has a median depth of merely $\sim100\Mpch$ ($z_\mrm\simeq0.04$, Huchra \etal 2011, in preparation\footnote{For current status see \texttt{http://tdc-www.cfa.harvard.edu/2mrs/}.}). On the other hand, the `parent' catalog of this survey, namely 2MASS XSC, reaches 3 times deeper and includes over 20 times more galaxies. Its redshift coverage, when matched with other surveys, such as SDSS \citep{SDSS} or 6dFGS \citep{6dF}, is non-uniform both on the sky and in depth (for the most recent compilation see \citealt{LH11}). For the purpose of our work, we have decided to sacrifice the advantages of weighting galaxies, possible for redshift surveys, obtaining instead an overwhelmingly greater number of sources and unprecedented depth of the survey with photometric data only. An additional motivation of using the dipole (\ref{eq:lim flux dipole}), constructed only with the us of fluxes and angular positions of individual galaxies, is the fact that it is free of any redshift-space distortions, and in particular of the rocket effect \citep{Kaiser}. The latter consists in the fact that the peculiar acceleration of the LG calculated using redshifts instead of real distances will differ from the actual LG acceleration due to a spurious contribution from the galaxies that are in the direction of the LG motion. Here, we do not use distances measured in redshift- nor in real space, and the only possible effect of that kind would be the anisotropy modulation in the distant galaxy distribution due to the aberration effect, which is however completely negligible for our sample \citep{Itoh}. The only stage at which the Kaiser effect comes into play is in the measurement of the luminosity function $\Phi(L)$ and consequently the luminosity density $j$. This is addressed in the relevant papers where $\Phi(L)$ is measured, see e.g.\ \cite{6dF_Fi}.

\ind The main interest of this paper is to examine the dipole (\ref{eq:lim flux dipole}) as a function of increasing depth of the 2MASS galaxy sample, as described in detail in the following chapters. This analysis will allow us to constrain the $\beta$ parameter from this dipole and consequently to measure the $\Omm$ parameter. Our method however is not to directly compare the peculiar velocity and acceleration of the Local Group; instead, we use the observed growth of the dipole to obtain these constraints.
\section{Data preparation}\label{Sec:Data} 

The Two Micron All Sky Survey (2MASS, \citealt{Skr}) is the first near-infrared survey of the whole sky (covering 99.998\% of the celestial sphere), and was performed in the period 1997--2001 in the $J\, (1.25\, \mu\mathrm{m})$, $H\, (1.65\, \mu\mathrm{m})$ and $K_s\, (2.16\, \mu\mathrm{m})$ bands, with the use of twin 1.3-m ground-based telescopes.  All the data from the survey are available through the NASA/IPAC Infrared Science Archive.\footnote{\texttt{http://irsa.ipac.caltech.edu/Missions/2mass.html}} The main outcome of this project are two photometric catalogs: of point sources (PSC), containing about 471 million objects, and of extended ones (XSC), with more than 1.6 million objects, mainly galaxies ($>98\%$) and some diffuse Galactic sources \citep{Jar04}. The XSC, which was used for the purpose of our analysis, is complete for sources brighter than $K_s\simeq13.5$ mag ($\sim2.7\,\mathrm{mJy}$) and resolved diameters larger than $\sim10$ -- $15''$. The near-infrared flux is particularly useful for the purpose of large-scale structure studies as it samples the old stellar population, and hence the bulk of stellar mass, and it is minimally affected by dust in the Galactic plane \citep{Jar04}. An additional advantage of using 2MASS data, especially in the context of calculating the flux dipole, for which \emph{apparent} magnitudes are used, is the global photometric uniformity of the catalog, which was enforced by nightly photometric calibration to an extensive set of standard star fields. On the other hand, as any survey, 2MASS is not perfect. It is biased against optically blue and low surface brightness galaxies, such as dwarfs, but sensitive to the early type, bulge-dominated ones. As the former have very small luminosities and masses, their possible underrepresentation in the catalog should not influence significantly our results.

The 2MASS photometry offers several types of `magnitudes' for extended objects, depending on the type of aperture used etc. Throughout the whole analysis we use the $20\,\mathrm{mag\slash{}sq.''}$ isophotal fiducial elliptical aperture magnitudes, which are defined as magnitudes inside the elliptical isophote corresponding to a surface brightness of $\mu_\mathrm{band}=20\,\mathrm{mag\slash{}sq.}''$. We prefer those to the Kron ones as the latter use large and noisy apertures, prone to contamination, resulting in systematic overestimation. Our choice is additionally supported by the considerations in the appendix of \cite{Kochanek}. However, we must remember to correct the values used by adding an offset of $\Delta=-0.2$~mag when converting to flux (cf.\ Section~\ref{Sec:Growth}), in order to compensate for the flux lost outside the aperture (typically $\sim$ 10\% -- 20\%, \citealt{LGA}). We have checked that this offset is roughly equal to the one between isophotal fiducial elliptical aperture magnitudes and the `total' ones, obtained from fit extrapolation. The magnitude correction by a constant factor certainly introduces some scatter in total flux estimates, as in reality it may depend on galaxy morphology. The latter is very hard to constrain from 2MASS data, we will thus treat this scatter as a systematic effect that needs to be included in the error budget. Our tests show that this error is of the order of a couple percent.

In order to prepare the data for our purposes, we have proceeded as follows. First of all, we applied the extinction correction from \cite{SFD}, by calling the procedure \texttt{dust\_getval}\footnote{\texttt{http://www.astro.princeton.edu/\textasciitilde{}schlegel/dust/dustpub/CodeC/README.C}}  for galactic coordinates of each of the objects. The procedure yielded values of \mbox{$E(B-V)$}, which were subtracted from the original magnitudes with appropriate multiplicative factors $R_V$ taken from \cite{CCM}: $0.902$ for $J$, $0.576$ for $H$ and $0.367$ for $K$. We performed the subtraction for objects with \mbox{$|b|>5\dgr$} due to the statement of \cite{SFD} that for \mbox{$|b|<5\dgr$} the predicted reddenings should not be trusted (the sources in the ZoA were not included in our catalog apart from several brightest; see end of this Section). Moreover, for some minor parts of the sky, the extinction correction gave unreliably high reddenings, which resulted in some objects becoming unrealistically bright (with negative magnitudes) and eventually deleted. At this stage, we have also identified and removed the following objects (with some found in more than one category):
\begin{itemize}
\item[--] artifacts: flag \texttt{cc\_flg=a} in the 2MASS XSC (122 objects);
\item[--] sources with \texttt{NULL} or unreliable $K$ magnitudes (as described above) (718 objects);
\item[--] non-extended sources: flag \texttt{vc=2} in the 2MASS XSC (7383 objects);
\item[--] Local Group galaxies, taken from the list of \cite{Lees} (31 objects);\footnote{Not all the objects from the Local Group were found in the database. These were some dwarf galaxies of low mass and near-IR luminosity, hidden behind the Galaxy or with surface brightness below the threshold of 2MASS.}
\item[--] Milky Way sources, identified by cross-correlation with a list of such 4454 objects, separately identified earlier in the 2MASS XSC.
\end{itemize}

The 2MASS Extended Source Catalog contains mainly galaxies; however, it is also comprised of Milky Way entities, such as stellar clusters, planetary nebulae, HII regions, young stellar objects and so on. In order to keep our analysis reliable, these objects had to be removed from the catalog. This was partially done for the 4454 sources mentioned above. However, owing to the size of the catalog, any further `manual' procedure of Galactic object removal was impossible and only a method based on some general properties could be applied. A useful one in this regard is the color, i.e.\ difference of magnitudes in two bands. In their analysis, \cite{Maller03} made a cross-correlation with galaxies spectroscopically confirmed from the Sloan Digital Sky Survey (SDSS) and excluded extended sources brighter than \mbox{$K=12$ mag} with colors \mbox{$J-K<0.75$} or \mbox{$J-K>1.4$}; at fainter magnitudes only those objects with \mbox{$J-K<0.5$} were removed. We have checked these conditions by examining the \mbox{$J-K$} distribution of galaxies in the 2MASS Redshift Survey (2MRS), which is a complete subset of the 2MASS XSC, containing all the galaxies with \mbox{$K <11.75$ mag}, with measured redshifts (more than 43,000 objects). Using the data from the Extragalactic Distance Database\footnote{\texttt{http://edd.ifa.hawaii.edu}; the complete 2MRS catalog is currently downloadable from \texttt{http://tdc-www.cfa.harvard.edu/2mrs/}.} \citep{EDD}, we have found that indeed galaxies are clustered around \mbox{$J-K=1$}; however, we have decided to alter the limits given by \cite{Maller03}. Analyzing additionally the distribution of XSC objects with \mbox{$K <13.5$ mag} and \mbox{$|b|>15\dgr$}, among which there are mainly non-Milky Way sources, apart from some molecular clouds (Ka{\l}u\.zny, private communication), we have decided to keep in our catalog those objects that have \mbox{$0.6<J-K<2.0$} (cf.\ also \citealt{NIRGMA}). We use this criterion for all sources, independently of magnitude, as we think that a differentiation as in \cite{Maller03} could lead to a bias in the sample. An additional eyeball verification of 100 brightest objects which pass this filter off the Galactic plane confirms that indeed all of them are galaxies and that only one galaxy with an extreme value of $J-K$ is removed by this procedure up to $K_s\sim7.5$ mag.

An important issue in the calculation of the clustering dipole is the Zone of Avoidance (ZoA), i.e.\ the region of the sky with small Galactic latitudes $b$, which obscures galaxies behind the Galactic plane and bulge. Although the Galactic extinction is much lower in the near infrared than in visible bands \citep{CCM} and this applies equally to the ZoA \citep{Jar00}, the 2MASS XSC is still incomplete near the Galactic equator, mainly due to high stellar density in this region of the sky \citep{KKJ}. For that reason, and owing to inapplicability of the extinction maps of \cite{SFD} for \mbox{$|b|<5\dgr$}, we have masked out the Galactic plane and bulge in the following way. For the shape of the mask we have chosen the one proposed in \cite{Erdogdu}, i.e.\ we have skipped all the objects with \mbox{$|b|<5\dgr$} (plane) and \mbox{$|b|<10\dgr$} for \mbox{$l<30\dgr$} or \mbox{$l>330\dgr$} (bulge). Then we have filled the resultant gap by cloning the adjacent strips, with mirror-like reflections: for instance, objects with $10\dgr<b<20\dgr$ were copied to the bulge by assigning \mbox{$b_\mathrm{new}:=20\dgr-b_\mathrm{old}$} and keeping other parameters unchanged (such as the longitude $l$ and magnitudes). An analogous procedure was used for the negative latitudes and for the Galactic plane. Such cloning has the advantage over random filling (considered both in \citealt{Maller03} and \citealt{Erdogdu}) that it extends the structures from above and below the ZoA; moreover, the only artificial discontinuity of the galaxy distribution created in this procedure is at the Galactic equator and at the edges of the box masking the bulge. We have also tried other masks and methods of filling the ZoA, and found no special importance for the results of the analysis presented here. This will be shortly addressed in Section~\ref{Sec:Growth}.

Once the ZoA has been masked and filled, we have added to our catalog several galaxies that were not present in the 2MASS XSC but could be found in the 2MASS Large Galaxy Atlas \citep[LGA,][]{LGA}. This atlas\footnote{Accessible through IRSA at \texttt{http://irsa.ipac.caltech.edu/applications/2MASS/LGA/}} contains the $\sim600$ largest galaxies as seen in the near-infrared, of which some 50 are not present in XSC or are located in the ZoA, $|b|<5\dgr$ (17 sources). Among the latter, three are of particular importance for the Local Group motion, namely Maffei 1 \& 2 and Circinus. We will discuss their influence on our results later in the text. Note that this addition does not spoil the photometric uniformity of the resulting sample because all the galaxies from the LGA present also in the XSC were assigned the magnitudes from the former catalog when the latter one was constructed.

In case of those LGA galaxies that were present in the ZoA, the \cite{SFD} maps are known to \textit{overestimate} the Galactic extinction by roughly 15\% \citep[e.g.][]{Schroder}, we have thus decreased the $E(B-V)$ by that amount there. The exceptions are Maffeis 1 \& 2, for which we used \textit{exact} values of extinction, given in \cite{Maffei2}, as well as Circinus with $E(B-V)=0.677$ (For \etal 2011, in preparation). Note however that apart from those three galaxies, which have extinction-corrected magnitudes below $K=5$~mag, all the remaining ones added from LGA are much fainter, at least by 2~mag, and possible misestimation of their extinction does not largely influence our analysis.

\section{Growth of the 2MASS dipole}\label{Sec:Growth}
In this section we will use the data prepared as described in Section~\ref{Sec:Data} to calculate the clustering dipole of the galaxies from the 2MASS XSC and analyze its growth. For that purpose we change units of $\bmd$ into $\kms$, defining the \emph{scaled dipole} as
\begin{equation}\label{eq:g_v} 
\tilde{\bmd} =\frac{8\,\pi}{3\, H_0\, \Omega_\mrm}\,\bmd =\frac{H_0}{b\,j}\sum_i^N S_i\hat{\bmr}_i\;,
\end{equation}
where we have used the Formula (\ref{eq:lim flux dipole}) for the flux dipole of the survey. 

The flux of each galaxy is calculated from its magnitude $m_i$ as
\begin{equation}\label{eq:S_i}
S_i=S_0\,10^{-0.4m_i}\,,
\end{equation}
where $S_0$ is the flux for a 0-magnitude object. As was already stated, we consider magnitudes in the  \mbox{$K_s$} band, which was the main (`target') band of the 2MASS survey (for simplicity of notation, we sometimes skip the `$s$' subscript). Then Eq.\ (\ref{eq:g_v}) takes on the form:
\begin{equation}\label{eq:g_v_Ks}
\tilde{\bmd}=\frac{H_0}{b\,j_K}\,S(0\,\mathrm{mag}) \sum_{i} 10^{-0.4(K _i+\mathrm{ZPO})} \hat{\mathbf{r}}_i\;,
\end{equation}
where \mbox{$S(0\,\mathrm{mag})=1.122\times10^{-10}\,\mathrm{W\,m^{-2}}$ $(\pm1.685\%)$} and the zero point offset $\mathrm{ZPO}=0.017\pm0.005$ \citep{CWM}. The $K_i$ magnitudes in Eq.\ (\ref{eq:g_v_Ks}) include also a negative offset of $\Delta=-0.2$ mag added following \cite{Kochanek} due to the underestimation of total fluxes by the isophotal magnitudes in the 2MASS XSC. The quantity $j_K$ is the luminosity density in the $K$ band.
It is obtained for example from the integral (\ref{eq:lum.dens}) using the luminosity function in this band. The value of $j_K$ has been estimated by many authors in the recent decade: \cite{Kochanek, Cole, Bell, Eke, 6dF_Fi, SLC}. We use here the one we consider the most reliable, calculated by \cite{6dF_Fi} from the luminosity function of more than 60,000 galaxies in the 6-degree Field Galaxy Survey (6dFGS, \citealt{6dF}): $j_K=(5.9\pm0.6) \times 10^8\, h\, \mathrm{L}^K  _\odot \, \mathrm{Mpc}^{-3}$, with $\mathrm{L}^K _\odot\simeq6.8\times10^{24}\,\mathrm{W}$ \citep{Rieke}.\footnote{$\mathrm{L}^K _\odot$ is the Solar luminosity in the $K$ band, calculated based on table 7 of \cite{Rieke},  \texttt{http://www.iop.org/EJ/article/1538-3881/135/6/2245/aj271287\_mrt7.txt}.} Note that as the luminosity density depends on the Hubble constant, the $h$ factor cancels out and Eq.\ (\ref{eq:g_v_Ks}) may be rewritten as 
\begin{equation}\label{eq:dim-less dipole}
\tilde{\bmd}=\frac{\mathcal{C}}{b} \, \sum_{K_\mathrm{min}}^{K_\mathrm{max}} 10^{-0.4K_i}\hat{\mathbf{r}}_i\;,
\end{equation}
where $\mathcal{C}\simeq 2620\,\kms$ for parameter values as given above. The limits of the summation are now the minimum and maximum $K$ magnitudes of galaxies in the sample. The lower limit in Eq.\ (\ref{eq:dim-less dipole}) is the magnitude of the brightest object in the catalog (excluding LG galaxies) and the upper one will be increased, as will be discussed later in the text (Section~\ref{Sec:Growth}). To retain the reliability of the analysis, we must note that the results remain trustworthy only up to the completeness limit of the catalog: the sample becomes incomplete for objects with \mbox{$K\gtrsim 13.5$ mag} \citep{Jar04}.

\begin{figure}
\centering\includegraphics[angle=270,width=\textwidth]{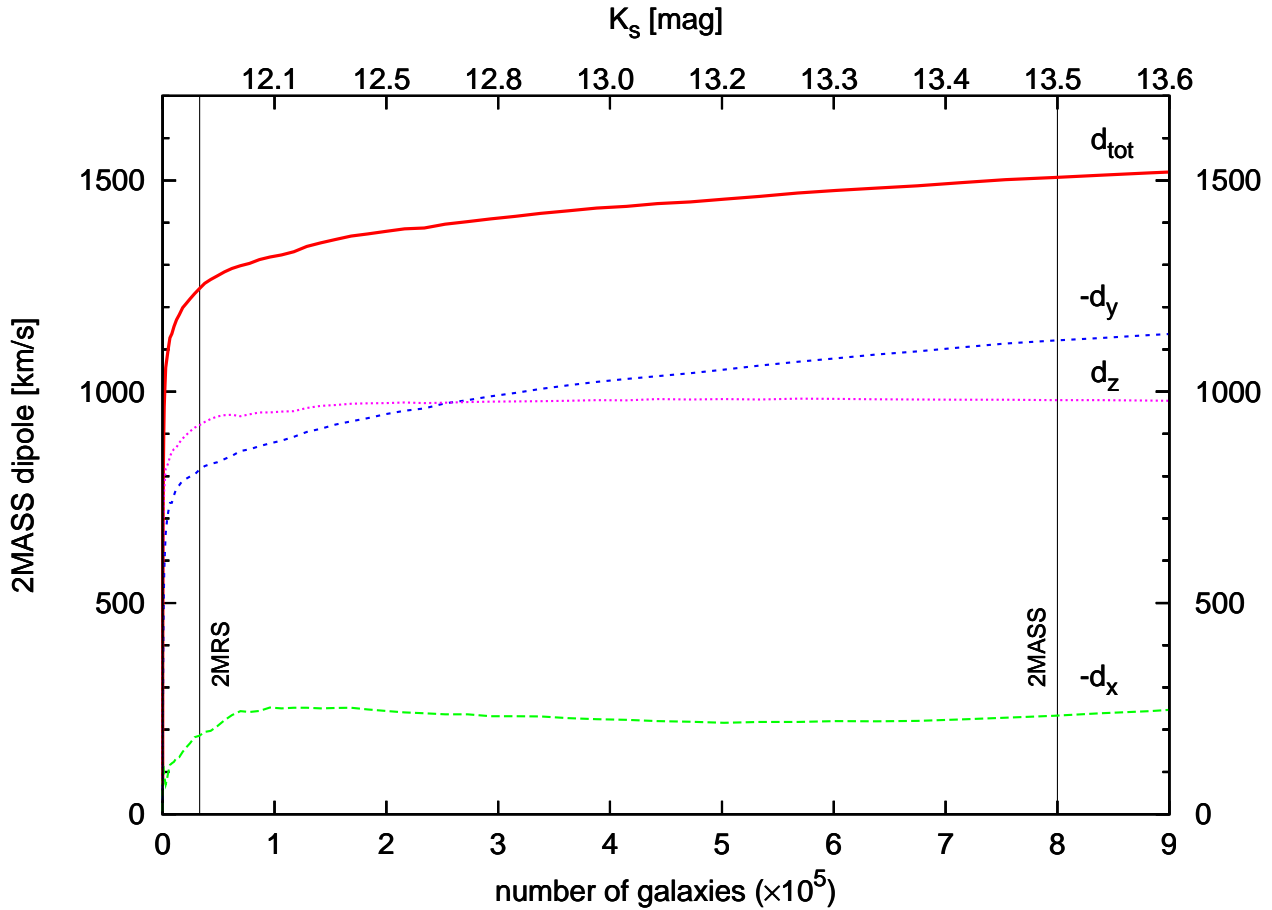}
\caption{Growth of the 2MASS clustering dipole as a function of the number of galaxies used for the calculation (bottom axis), ordered by their $K_s$ magnitudes (top axis). The thick red line is the amplitude of the dipole; the thin dotted and dashed lines (green, blue and magenta) are the Cartesian components (in Galactic coordinates). Two vertical lines illustrate the limits of the 2MRS $K_s\leq11.25$~mag sample and completeness of the 2MASS XSC ($K_s=13.5$~mag). Such data presentation could suggest that the 2MASS dipole has converged within sample limits, which is \textit{not} the case.}
\label{Fig:growth.N}
\end{figure}
\begin{figure}
\centering\includegraphics[angle=270,width=\textwidth]{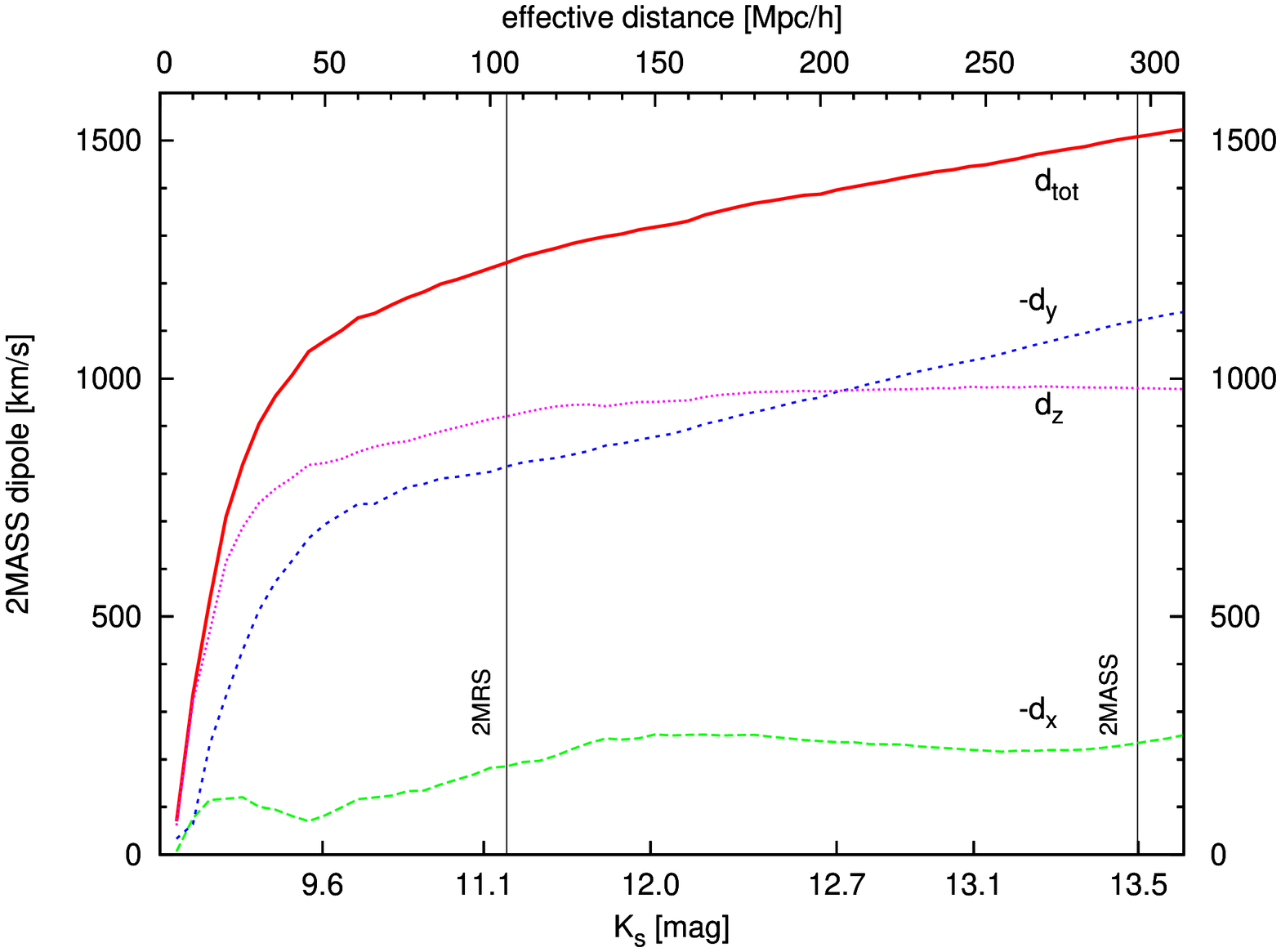}
\caption{Growth of the 2MASS clustering dipole as a function of increased maximum $K_s$ magnitude of the subsample (bottom axis). Corresponding  effective distance is given at the top. The thick red line is the amplitude of the dipole; the thin dotted and dashed lines (green, blue and magenta) are the Cartesian components (in Galactic coordinates). Two vertical lines illustrate the limits of the 2MRS $K_s\leq11.25$~mag sample and completeness of the 2MASS XSC ($K_s=13.5$~mag). The lack of convergence of the dipole is clearly visible. Note also the steady growth of the amplitude as well as of the Galactic $y$ component of the dipole.}
\label{Fig:growth.r}
\end{figure}
The growth of the dipole was calculated by incrementing $K_\mathrm{max}$ in the sum given in Eq.\ (\ref{eq:dim-less dipole}). Results are illustrated in Figures~\ref{Fig:growth.N} and \ref{Fig:growth.r}. Figure~\ref{Fig:growth.N} shows the growth of the dipole as a function of the number of galaxies used for the calculation, together with Galactic Cartesian components of the acceleration ($g_x$ points towards the Galactic center, $l=0\dgr$, $b=0\dgr$; $g_y$ is perpendicular to it in the Galactic plane; $g_z$ is perpendicular to the Galactic plane). The similarity of this plot to figure~1 of \cite{Maller03} is intentional and allows us to draw the conclusion that although our data analysis was slightly different to that of \cite{Maller03} (altered shape and filling of the ZoA, different removal of Galactic objects), the two approaches give qualitatively the same results. However, we want to stress that such a presentation of the data may be misleading. Owing to local inhomogeneities, a linear scale in the number of galaxies on the $x$-axis does not have to be related in a simple manner to a scale expressed in cut-off magnitudes nor in distances and `compresses' the left-hand side of the plot, while `stretching' the right-hand one. This results in a suggestion of convergence of the dipole, which was actually concluded by \cite{Maller03}. However, what we actually need in order to draw proper conclusions about the growth and possible convergence of the dipole, is a linear scale \emph{in distance} on the abscissa.

\ind Neither distances, nor even redshifts (including photometric ones) are currently measured for the whole 2MASS XSC (although some attempts are being made regarding photo-$z$'s, see \citealt{Jar04} and \citealt{FraPea}). We thus need to deduce effective distances of galaxies from their fluxes, with the use of the luminosity function (LF) in the $K$ band. If all the galaxies had the same luminosity, say $L_*$, the relation between the observed flux $S$ and distance $r$  would be straightforward: $r=\sqrt{L_*\slash 4\pi S}$. However, galaxies have different morphologies, masses and luminosities, and obviously their LF is not a Dirac's delta (on the contrary, it is very broad). Therefore an estimated distance of a galaxy with a given flux must have a scatter. Constructing the estimator, a first choice could be the conditional mean, i.e.\ the expectation value for $r$ given $S$. Instead, we think that it is better to choose the conditional \emph{median} for $r_\mathrm{eff}$ (a median value of distance given the flux). We consider it being more adequate to our problem: the same number of galaxies with a given flux have distances smaller and greater than the median.

Detailed calculations, presented in Appendix \ref{App:r_eff} (see also \citealt{Pe93}), show that for the $K$ band LF as given by \cite{6dF_Fi}, i.e.\ a Schechter function \citep{Sche} with $M_*=-23.83+5\log h\pm0.03$ and $\alpha=-1.16\pm0.04$, this effective distance for the magnitude $K$ equals to
\begin{equation}\label{eq:r_eff}
r_\mathrm{eff}\simeq0.59\times10^{0.2K}\Mpch \;.
\end{equation}
This proxy of distance is used in Figure~\ref{Fig:growth.r}, which differs from Figure~\ref{Fig:growth.N} by a different scaling of the $x$-axis. The growth of the clustering dipole up to the completeness limit of the 2MASS XSC is now evident. Additionally, note that the growth has an essentially constant slope for $r_\mathrm{eff}>150\Mpch$, i.e.\ $K_s>12$ mag.

An interesting feature is the behavior of the Galactic Cartesian components of the dipole. The $x$ and $z$ ones are virtually constant for $r_\mathrm{eff}>150\Mpch$; however, the $y$ component still grows even at the limit of the catalog, similarly as does the total amplitude. This could point to some systematic effect, related to masking and filling of the Zone of Avoidance. We have however checked that the same qualitative behavior of the three components is observed for different shapes of the mask and the way it is filled; what is more, the effect exists even if we calculate the dipole having removed from the catalog all the galaxies with $|b|<10\dgr$ (leaving the resulting strip completely devoid of galaxies). We have also observed that adding to the sample the LGA galaxies that were not present in our catalog has virtually no influence on the amplitude of the dipole and only slightly changes the values of particular components (see also below).

\section{Misalignment angle}\label{Subsec:Angle}
\begin{figure}
\centering\includegraphics[angle=270,width=\textwidth]{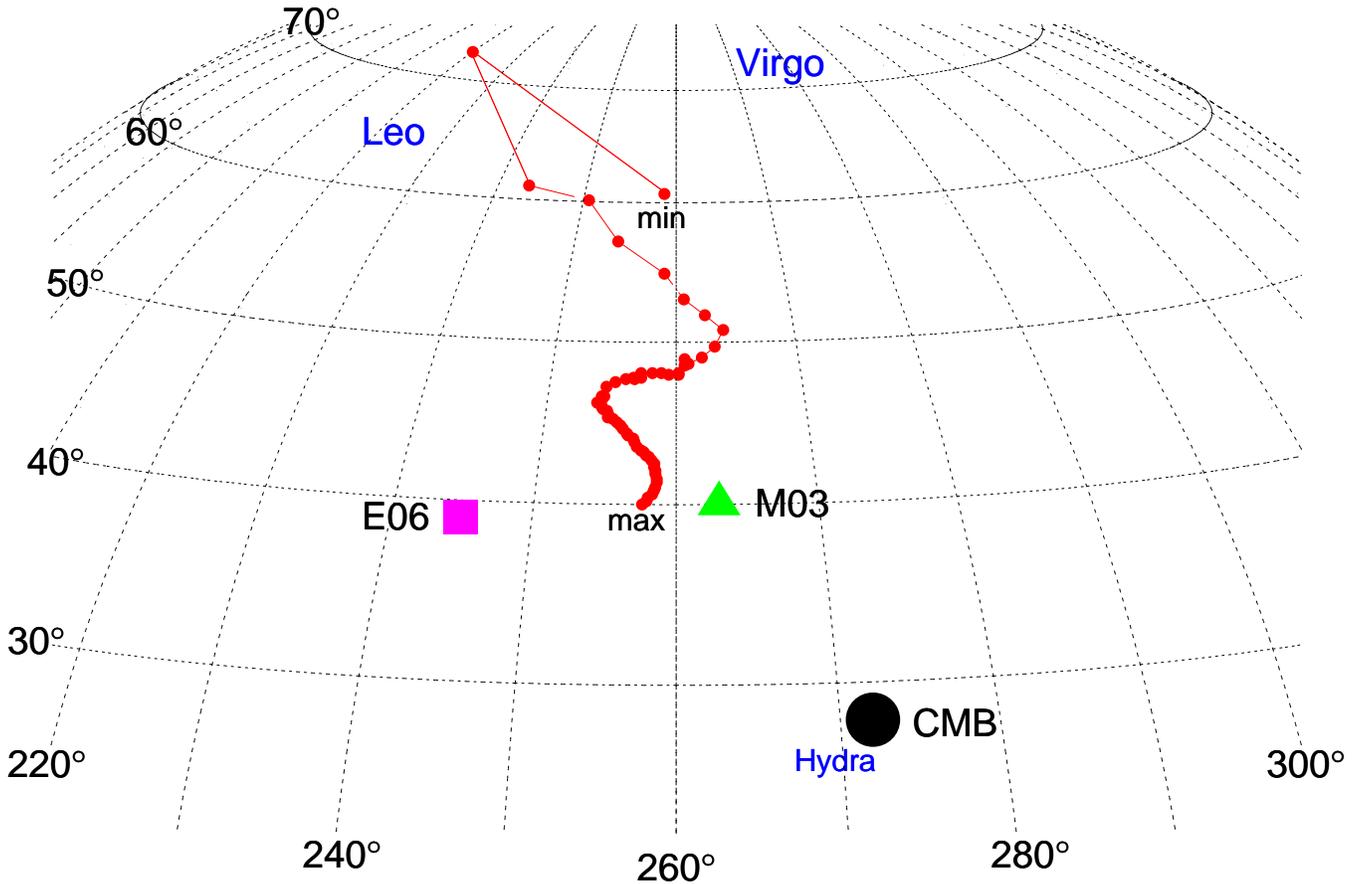}
\caption{Position of the 2MASS clustering dipole on the sky as a function of increased depth of the sample. The grid shows Galactic coordinates: longitude $l$ (labels on bottom) and latitude $b$ (left). The min/max labels refer to minimum and maximum depth of the subsample for which the dipole was calculated, respectively $5\Mpch$ and $310\Mpch$. The black disc marked by `CMB' is the direction of the peculiar velocity of the Local Group ($l=272\dgr$, $b=28\dgr$). The green triangle labeled `M03' shows the 2MASS clustering dipole of Maller \etal (2003): $l=263\dgr$, $b=40\dgr$ (with cloning in the ZoA). The magenta square marked by `E06' is the direction of the flux-weighted 2MRS dipole in the CMB frame of Erdo{\u g}du \etal (2005): $l=245\dgr$, $b=39\dgr$. Blue labels indicate important structures in the local Universe: the Virgo Cluster ($D\simeq17\Mpc$), the Hydra Cluster ($D\simeq47\Mpc$) and the Leo Supercluster ($z\gtrsim0.031$).}
\label{Fig:on.the.sky}
\end{figure}
\begin{figure}
\centering\includegraphics[angle=270,width=\textwidth]{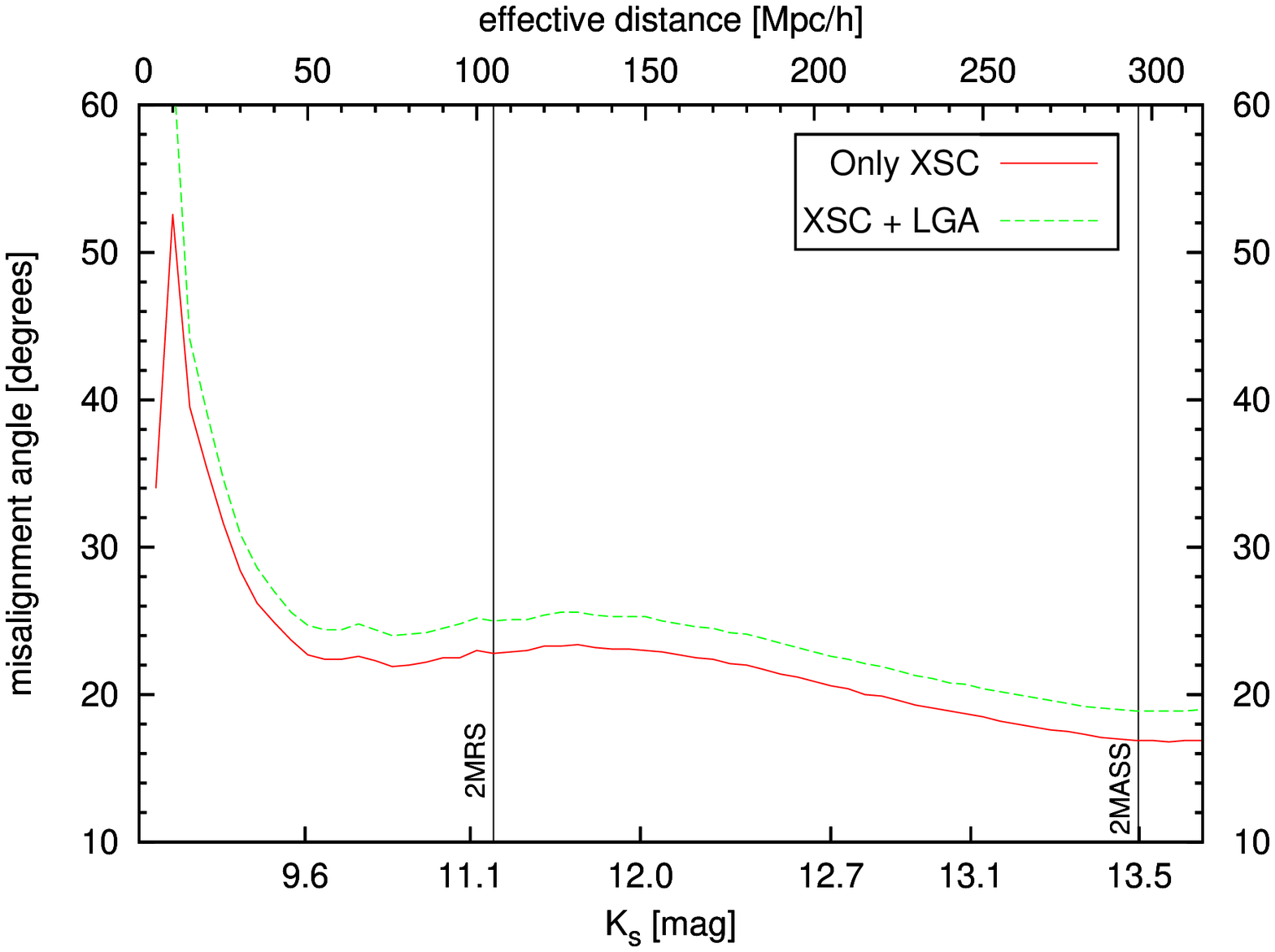}
\caption{Misalignment angle between the 2MASS clustering dipole and the peculiar velocity of the Local Group as a function of increased maximum $K_s$ magnitude of the subsample (bottom axis). The corresponding effective distance is given at the top. Two vertical lines illustrate the limits of the 2MRS $K_s\leq11.25$~mag sample and completeness of the 2MASS XSC ($K_s=13.5$ mag). The solid and dashed lines are shown to illustrate the effect of adding the Large Galaxy Atlas  to the sample.}
\label{Fig:misal}
\end{figure}
The linear theory Relation (\ref{eq:v.and.g}) between peculiar velocity and acceleration of the Local Group predicts that the two vectors should be parallel. In reality, a non-zero \textit{misalignment angle} between them is expected, due to various reasons, such as the scatter in the mass-to-light ratio (e.g.\ \citealt{Cr07}) or stochasticity in the non-linear relation between the velocity and acceleration of the LG \citep{B99, CCK01, CCBCC}. Observed values of this angle are usually of the order of 10 -- 20 degrees \citep{Strauss, Schmoldt, Maller03, Erdogdu} and we have obtained similar results, which confirms the validity of the linear approximation for the scales of interest. Figure~\ref{Fig:on.the.sky} shows how the dipole direction on the sky changes as the sample depth increases, from `min', i.e.\ $5\Mpch$, to `max', equal to $310\Mpch$. The black disc shows the CMB dipole. For comparison, we present also two other results of 2MASS data analysis: 2MASS dipole with `cloning' the ZoA, from \cite{Maller03} (green triangle labeled M03) and the flux weighted 2MRS dipole in the CMB frame of \cite{Erdogdu} (magenta square, E06). We increase the Galactic longitude $l$ from left to right for easier comparison with relevant figures in these two papers (respectively fig.\ 2 of M03 and fig.\ 7 of E06).

In Figure~\ref{Fig:misal} we plot the misalignment angle with respect to the CMB dipole direction, as a function of growing depth of the sample. The two lines illustrate the effect of adding the LGA galaxies to the catalog, and in particular of the three bright galaxies located in the ZoA (Maffeis 1 \& 2 and Circinus). A specific `tug-of-war' between these structures, located on almost opposite sides of the Galactic plane, results in a general raise of the angle by $\sim1\degr$. As can be also seen, the brightest (and presumably the closest) galaxies have the most influence on the value of the angle. Note also that already for a distance as small as $50\Mpch$ the misalignment angle reaches a value of $\sim20\dgr$ ($\sim0.35$~rad), with a minimum of $16.8\dgr$ for $r_\mathrm{eff}=305\Mpch$ (without the LGA galaxies).

One can think of several reasons for this misalignment. The first possibility are influences from extremely large scales. We cannot address this issue directly here, note only that the claims of \cite{WFH09} or \cite{KABK} of large-scale bulk flows with values much larger than expected in $\Lambda$CDM have been recently put in doubt by \cite{NBD11a,ND11} or \cite{Osborne}; the debate is thus far from being settled. The second option could be unsuspected structures in the Zone of Avoidance. This issue was partly addressed in \cite{BC10}, where the influence of one of such partially obscured objects -- the Local Void (LV) -- was analyzed. As was shown, the improper filling of the ZoA in the LV region cannot bias the angle by more than $~1\degr$. Additionally, already in 2005 it was quite certain that thanks to surveys in various wavelengths (such as HI or X rays) there should be no unknown nearby large-scale structures hidden behind the Milky Way \citep[e.g.][]{FaLa05}. Finally, such a big value of the misalignment angle can be due to improper accounting of very local influences; we have tested this possibility by adding the LGA galaxies to our sample.

Note however that our statistical assumption that the flux is a good proxy of the gravitational force, which may be appropriate for the whole sample, can fail for individual galaxies. In particular, we have used a constant mass-to-light ratio for all galaxies, which is probably valid when averaged over many of them, especially since we use the $K_s$-band luminosity, which is known to be a better tracer of stellar mass than optical and other NIR wavebands. Still, the $M/L_K$ ratio should vary somewhat with morphology and luminosity. For galaxies with smallest magnitudes these variations may significantly affect their contribution to the direction of the clustering dipole. The same was noted by \cite{Maller03}, who also concluded that the $16\degr$ difference between their 2MASS dipole and the CMB one was caused by the brightest galaxies. We obtain a similar result here and interpret it as due to a specific non-linear effect, the \textit{shot noise}, which translates into the fact that just a few sources with the largest fluxes have a large influence on the misalignment.

The high significance of the most luminous (in terms of the observed flux) galaxies for the direction of the dipole confirms theoretical predictions of \cite{CCBCC}: the misalignment angle could be lowered if the observational window used was the optimal one, which would be the case if we \textit{removed} those brightest galaxies from the sample (as was also tried both by \citealt{Maller03} and \citealt{Erdogdu}). We plan to address this issue in a forthcoming paper concerning the maximum-likelihood estimation of $\beta$ from the 2MASS dipole.

\section{Discussion}\label{Sec:Discussion}
Our results showing that the clustering dipole of 2MASS galaxies still grows even at the completeness limit of the sample, taken at face value, are consistent with the results of some other authors, who used various datasets and methods: redshift survey of \emph{IRAS} galaxies, \cite{Strauss}; \emph{IRAS} PSC\emph{z}, \cite{Schmoldt} and reanalysis by \cite{BP06}; X-ray selected clusters, \cite{KE06}; reconstructed velocity field of 2MRS, \cite{Lav10}. On the other hand, they contradict claims of convergence at scales even as small as 60 -- 100$\Mpch$: optical sample of \cite{Hudson}; redshift sample of Abell/ACO clusters, \cite{Tini}; \emph{IRAS} PSC\emph{z}, \cite{RR2000}; \emph{IRAS} PSC\emph{z} and BTP, \cite{DMSa}; 2MRS, \cite{Erdogdu}. Our analysis also suggests a different interpretation of the results of \cite{Maller03} --- data presentation as in our Figure~\ref{Fig:growth.r} instead of figure~1 therein would possibly point to similar lack of convergence. On the other hand, in order to ba able to \textit{directly} compare our results with those of \cite{Erdogdu}, we would have to apply the same weighting as was done for the 2MRS sample, namely by the inverse of the flux-weighted selection function. We are unable to do it, not knowing distances nor redshifts for the whole sample. 

\ind Apart from galaxy weighting, the discrepancies between the above listed results most probably stem from the different nature of catalogs and methods used for the calculation, and in particular may be due to distinct \emph{observational windows}. Such a window for a given survey describes the sample: it may be interpreted as a filter (in real or Fourier space) through which we observe the Universe. Knowledge of the observational windows, necessary to correctly confront results as those given above, is also essential if we want to make comparisons with theoretical expectations.

\subsection{Theoretical framework}\label{Subsec:Theory}
We would now like to check if the behavior of the 2MASS flux dipole is consistent with the predictions of the currently favored cosmological model, namely Lambda-Cold-Dark-Matter ($\Lambda$CDM). We start by presenting the theoretical framework for such a comparison. It was first derived in the context of then-popular models like cold-dark-matter and isocurvature baryon, as described in detail in two classic papers: \cite{JVW} and \cite{LKH}. More recently, this approach was taken by \cite{Lav10}, who reconstructed the local peculiar velocity field (up to $\sim150\Mpch$), applying the data from the 2MRS. The basic quantity for these comparisons is the joint probability distribution function for $\bmv$ and $\bmg$, assumed to be a multivariate Gaussian.

In our case, we want to find the expectation value for the amplitude of the acceleration of the Local Group \emph{knowing} its peculiar velocity. The latter equals to $v_\mathrm{CMB}=622\pm35\,\mathrm{km\slash s}$ in the direction $(l,b)=(272\dgr\pm3\dgr\!,\,28\dgr\pm5\dgr)$. In our analysis we do not use the directional constraint: the misalignment angle is integrated out \citep[cf.][]{JVW}. In the forthcoming paper, concerning the maximum-likelihood determination of $\beta$ from 2MASS XSC, we \textit{will} use the fact that the misalignment angle is known. Here, the theoretical prediction gives us the conditional velocity, $\bmv_\mrc$ (in units of $\kms$), which is related to the acceleration via the linear-theory relation (\ref{eq:v_s}). The relevant formula for its amplitude has the form given by eq.\ (8a) of \cite{JVW} (note a typo therein):
\begin{eqnarray}\label{eq:v_cond}
v_\mrc & \equiv & \langle v_m\vert v_\mathrm{CMB}\rangle =  \nonumber \\
&&=\sigma_m \frac{1-\rho^2+\rho^2 u^2}{\rho u}\,\mathrm{erf}\left\{\frac{\rho u}{\left[2(1-\rho^2)\right]^{1\slash 2}}\right\}+ \sigma_m\sqrt{\frac{2}{\pi}(1-\rho^2)}\,\exp\left[-\frac{\rho^2 u^2}{2(1-\rho^2)}\right]\;, 
\end{eqnarray}
where we have used the following quantities:
\begin{itemize}

\item
$\bmv_m$ is the velocity induced by a single realization of the density field given the assumed power spectrum
\item[]
\begin{equation}\label{eq:v_m}
\bmv_m=\frac{H_0\, f(\Omm)}{4\,\pi}\int{\delta(\bmr)\,\frac{\bmr}{r^3}\,W(\bmr)\,\de^3 \bmr}\;,
\end{equation}
\item[]with $\delta(\bmr)$ and $W(\bmr)$ being respectively the density field and the window of the survey. The latter is included to mimic the velocity measured from a given survey;
\item
$\sigma^2_m$ is the predicted (ensemble average) variance of the velocity measured from the survey, given by
\item[]
\begin{equation}\label{eq:sigma_m}
\sigma^2_m=\frac{H_0^2\, f^2(\Omm)}{6\,\pi^2}\int\limits_0^\infty{P(k) \tilde{w}^2(k)\,\de k}\;,
\end{equation}
\item[]with $P(k)\equiv \langle\vert\delta_\bmk\vert^2\rangle$ being the power spectrum of density fluctuations, $\bmk$ the wavevector and $\tilde{w}(k)$ the observational window in Fourier space;
\item $\bmu\equiv\bmv_m\slash\sigma_m\;;$
\item
$\rho$ is the correlation coefficient of $v_m$ and $v_\mathrm{CMB}$
\item[]
\begin{equation}\label{eq:rho}
\rho=\frac{H_0^2}{6\,\pi^2\,\sigma\,\sigma_m}\int\limits_0^\infty{P(k) \tilde{w}(k)\,\de k}\;,
\end{equation}
\item[]where $\sigma$ is given by Eq.\ (\ref{eq:sigma_m}) with $\tilde{w}(k)\equiv 1$.

\end{itemize}
The second moment of the conditional velocity is given by eq.\ (8b) of \cite{JVW}:
\begin{equation}
\mu_\mrc\equiv\langle v^2_m\vert v_\mathrm{CMB} \rangle=3\sigma_m^2(1-\rho^2)+\rho^2 u^2 \sigma_m^2\;,
\end{equation}
hence the variance is equal to
\begin{equation}\label{eq:sigma_cond}
\sigma^2_\mrc=\mu_\mrc-v^2_\mrc\;.
\end{equation}
The presence of the factor $f(\Omm)$ in Eqs.\ (\ref{eq:v_m})--(\ref{eq:sigma_m}) and the fact that we also have to include biasing $b$ mean that in order to properly compare the conditional velocity $v_\mrc$ given by Eq.\ (\ref{eq:v_cond}) with the dipole $d$ measured from the data, Eq.\ (\ref{eq:g_v}), the latter has to be rescaled, using a best-fit $\beta$ parameter, as in Eq. (\ref{eq:v_s}). In fact, this rescaling enables us to estimate $\beta$ in a straightforward way; we come back to this point a few paragraphs later.

\subsection{Observational window}\label{Subsec:Windows}
In this subsection we discuss the proper observational window for our measurement. We start by noting that when calculating the peculiar gravitational acceleration of the Local Group from the dipole of the all-sky galaxy distribution, in general there are two schemes to postulate a relation between this distribution and that of the underlying mass (see e.g.\ \citealt{Erdogdu}). The first scheme, called \emph{number weighting}, assumes that the mass is distributed in the Universe as a continuous density field, which is sampled by galaxies in a Poissonian way. The second prescription, namely \emph{flux weighting}, uses the assumption that all the mass in the Universe is locked to the mass of the halos of luminous galaxies. The 2MASS dipole as calculated in Sec.\ \ref{Sec:Growth} is of the flux-weighted type. 

The observational window $W(r)$ of a flux-limited survey is just its selection function $\varphi(r)$. In the number-weighted scheme, this function measures what fraction of galaxies located in the distance interval $(r,\,r+\de r)$ are included in the survey:
\begin{equation}
\varphi(r)=\frac{\int\limits_{4\pi r^2 S_\mathrm{min}}^\infty \Phi(L) \de L}{\int\limits_0^\infty \Phi(L) \de L}\;,
\end{equation}
where $\Phi(L)$ is the luminosity function and $S_\mathrm{min}$ is the limiting flux of the survey (e.g.\ \citealt{DH82, YSDH}). In our case of the flux-weighted dipole, as given by Eq. (\ref{eq:lim flux dipole}), the appropriate window is the flux-weighted selection function, $\Psi(r)$, defined in such a way that $1-\Psi(r)$ is the percentage of light from a distance $r$ which is not visible for the survey \citep{Erdogdu, CCBCC}. It is given by
\begin{equation}\label{eq:Psi_r}
\Psi(r)=\frac{\int_{L_\mathrm{min}}^\infty L\, \Phi(L) \de L}{\int_0^\infty L\, \Phi(L) \de L}\;,
\end{equation}
where $L_\mathrm{min}=4\pi r^2 S_\mathrm{min}$. The window $W_S(r)\equiv\Psi(r)$, where the subscript $S$ emphasizes the dependence of the window on the flux limit of the catalog, smoothly decreases to zero with increased distance from the observer. For a detailed discussion see \cite{CCBCC}.

On the other hand, `ideal' surveys are distance- (or volume-) limited: in this case there is no loss of signal up to the limiting distance $R$. Then the window function is simply unity for $r<R$ and zero otherwise. It has thus a form of a Heaviside step function; for a spherical survey we have $W_R(r) = \Theta_{\mathrm H}(R - r)$ (e.g.\ \citealt{JVW}). In order to better reconstruct the dipole of the galaxy distribution from a flux-limited catalog, volume-weighting of the survey is commonly mimicked by weighting individual galaxies by the inverse of the selection function, at the expense of increasing shot noise from large scales \citep{Strauss} and at a price of a possibly large rocket effect \citep{Kaiser}. The top hat is then the relevant window. In our case however, as we are dealing with angular (photometric) data only, we cannot estimate the selection function even in redshift space. Therefore, we do not weight galaxies in the sample (or in other words, we assign unit weights to them) and the relevant window function is given by (\ref{eq:Psi_r}).

The Fourier-space counterpart of any observational window, $\tilde{w}(k)$, is obtained from the real-space one as \citep{KaLa, LKH, JVW}
\begin{equation}
\tilde{w}(k)\equiv k\int_0^\infty{W(r)\,j_1(kr)\,\de r}\,,
\end{equation}
where $j_1$ is the spherical Bessel function of the first kind of order 1. Note that this window is \emph{not} the Fourier transform of $W(r)$. In our case of $W(r)$ given by (\ref{eq:Psi_r}), the Fourier-space window reads (eq.\ 59 of \citealt{CCBCC})
\begin{equation}\label{eq:2MASS window}
\tilde{w}_s(k)=1-\frac{8\pi S_\mathrm{min}}{k\int_0^\infty{L\,\Phi(L)\,\de L}}\int\limits_0^\infty{\sin(kr)L_\mathrm{min}\,\Phi(L_\mathrm{min})\,\de r}\;,
\end{equation}
where $S_\mathrm{min}$ and $L_\mathrm{min}$ are as above. The window $\tilde{w}_s$ is a function of both the wavenumber $k$ and the minimum flux $S_\mathrm{min}$ (hence the subscript $s$). On the other hand, it does \emph{not} depend on the distance, as the latter is integrated out, together with absolute luminosities (with the use of the luminosity function). This means that when comparing observations to theoretical expectations, we never use the effective distance (\ref{eq:r_eff}) for the calculations of the conditional velocity $v_\mrc$ (\ref{eq:v_cond}) and its variance $\sigma_\mrc$ (\ref{eq:sigma_cond}). Our observable is the minimum flux of the sample, related to maximum magnitude via $S_\mathrm{min}=S_0\,10^{-0.4K_\mathrm{max}}$. For clarity however, in the relevant plots we prefer to show the results in terms of a linear scaling in the effective distance. 

\begin{figure}
\centering\includegraphics[width=\textwidth]{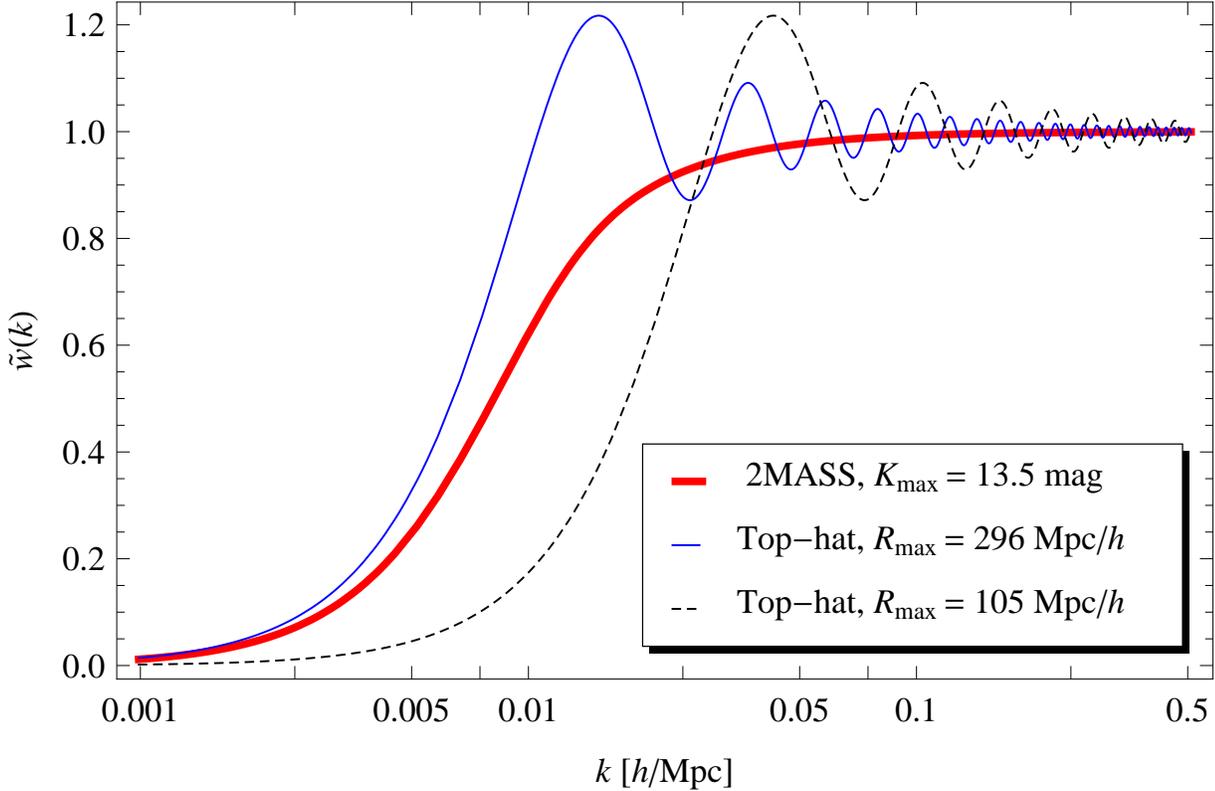}
\caption{Comparison of three observational windows in the Fourier space: the one of the 2MASS flux-limited survey (thick red line) for a maximum magnitude of $13.5$ mag and two for distance-limited surveys (top-hat) with $R_\mathrm{max}=296\Mpch$ (blue thin line) and $R_\mathrm{max}=105\Mpch$ (black dashed line). The former radius is the effective distance of galaxies at the limit of the 2MASS XSC; the latter corresponds to $K=11.25$ mag, the limit of the 2MRS.}
\label{Fig:windows}
\end{figure}

In Figure \ref{Fig:windows} we plot the 2MASS flux-limited window with $K_\mathrm{max}=13.5$ mag and compare it with two top-hat windows for distance-limited surveys: one with $R_\mathrm{max}=296\Mpch$ and the other with $R_\mathrm{max}=105\Mpch$, which are respectively the effective distances of galaxies at the limit of the 2MASS and 2MRS $K_\mathrm{max}=11.25$~mag sample. The Fourier form of the top-hat window is \citep{JVW}
\begin{equation}
\tilde{w}_\mathrm{th}(k)=1-j_0(k R_\mathrm{max})\;
\end{equation}
with $j_0\equiv\sin x\slash x$ being the spherical Bessel function of the first kind of order 0. The oscillating behavior of this window in the Fourier space is a result of the rapid truncation in the real space. As we can see, the top-hat window with $R_\mathrm{max}=296\Mpch$, which asymptotically behaves in the same way as the 2MASS one with $K_\mathrm{max}=13.5$ mag, passes clearly more large-scale signal ($k\lesssim0.02\hMpc$). This will result in much faster convergence of the dipole measured through the top-hat window than through the 2MASS flux-weighted one (see below). On the other hand, the top-hat window with a cut-off equivalent to the limit of 2MRS at $105\Mpch$ blocks almost all the signal already for $k\lesssim0.01\hMpc$, as expected.

\subsection{Comparison with theoretical predictions}\label{Subsec:Compare}
We start by reminding that our whole analysis is performed within the linear theory. For that reason, we do not include non-linear effects in the correlation coefficient $\rho$ (Eq.\ \ref{eq:rho}). Mathematically, this means that we set to unity two functions: the ratio of power spectra of density and velocity divergence, $\mathcal{R}(k)$, and the coherence function, $C(k)$. For non-linear analysis of the former, see \cite{CC04}. The latter, introduced by \cite{Strauss} (originally called the \emph{decoherence} function), can be interpreted as the cross-correlation coefficient of Fourier modes of gravity and velocity fields and was discussed in detail by \cite{CC02}.

Our goal here is to compare the observational data to the expectations of the $\Lambda$CDM concordance model. Hence the power spectrum used in Eqs.\ (\ref{eq:sigma_m}) and (\ref{eq:rho}) is the spectrum of cold dark matter, with baryon effects included, as given by \cite{EH98}. In this framework, $P(k)\propto k^{n_s} T^2(k)$, where $n_s$ is the spectral index of primordial fluctuations and the transfer function $T(k)$ depends on the parameters $h$, $\Omm$, $\Omb$ (density parameter of baryons) and $\sigma_8$ (present value of root-mean-square density contrast of matter fluctuations within a sphere of $8\Mpch$). In our calculations we use the following set of parameters obtained from \emph{WMAP} seven-year observations \citep{Larson}: $h=0.71$, $\Omm h^2=0.1335$, $\Omb h^2=0.02258$, $n_s=0.963$ and $\sigma_8=0.801$.

\begin{figure}
\centering\includegraphics[angle=270,width=\textwidth]{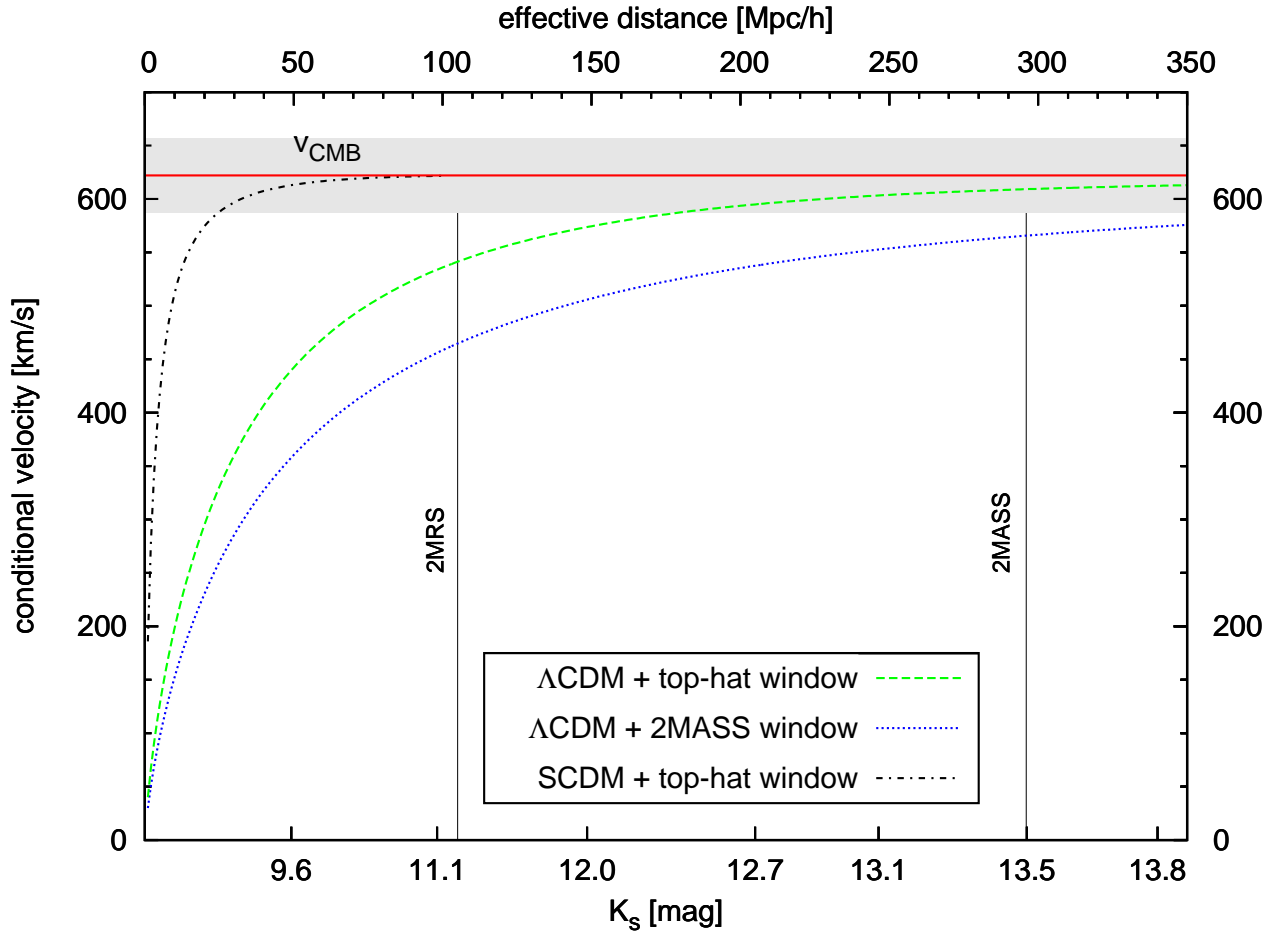}
\caption{Theoretically predicted growth of the conditional velocity of the Local Group for known $v_\mathrm{CMB}$ using two different observational windows: the one of 2MASS (blue dotted line) and top-hat (green dashed line), both calculated for the $\Lambda$CDM as the underlying cosmological model. The horizontal solid line is the observed velocity of the Local Group with respect to the CMB. For comparison, the prediction for the currently disfavored Standard CDM model with the top-hat window is shown with the black dot-dashed line.}
\label{Fig:predicted}
\end{figure}
\ind{}Before applying the observational data, in Figure~\ref{Fig:predicted} we compare the expected growth of the conditional velocity for the 2MASS window and for the top-hat case, both calculated for the $\Lambda$CDM model parameters. As we can see, for such a flux-limited survey as 2MASS, the predicted velocity is far from converging to the limit of $v_\mathrm{CMB}=622\,\mathrm{km\slash s}$ (horizontal line) even for $r_\mathrm{eff}\simeq350\Mpch$. In the top-hat case, the expected convergence is much faster, as could have been deduced already from the comparison of the windows, presented in Fig.\ \ref{Fig:windows}. However, it should be noted that even for all-sky catalogs that include redshifts, for which we can effectively model the dataset with a top-hat window (like in the case of the 2MRS), the convergence of the dipole is not likely before some $200\Mpch$, opposite to the results of \cite{Erdogdu}, where it is claimed that the contribution from structure beyond $6000\,\kms\,(=60\Mpch)$ is negligible. Figure~\ref{Fig:predicted} shows that for the latter distance, the conditional velocity for the top-hat window has reached less than $75\%$ of its final value. On the other hand, for the 2MASS window, $v_c$ attains $0.95\,v_\mathrm{CMB}$ no sooner than for $r\simeq470\Mpch$ ($\sim14.5$ mag in the $K_s$ band), far beyond the completeness of the 2MASS XSC. As an extreme case, in Figure~\ref{Fig:predicted} we additionally plot the prediction for the once-popular `Standard CDM' model, currently strongly disfavored by observations ($h=0.5$, $\Omm=1$, $\Omb=0.05$, $n_s=1$, $\sigma_8=1$), with the top-hat window. Note how fast convergence would be expected in this case.
\begin{figure}
\centering\includegraphics[angle=270,width=\textwidth]{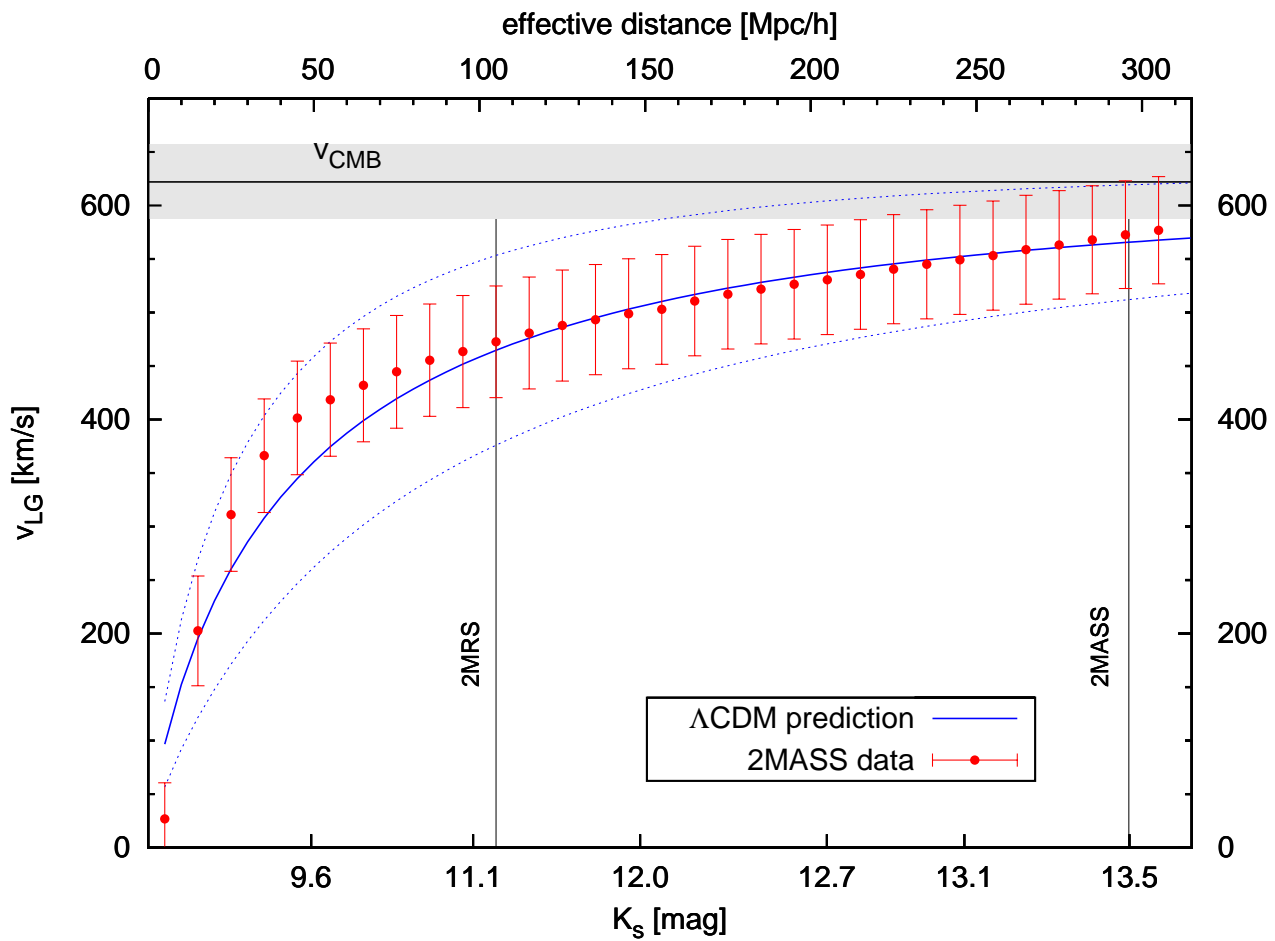}
\caption{Growth of the 2MASS clustering dipole (red points with 1$\sigma$ errorbars), compared with the theoretical expectation for conditional LG velocity in the $\Lambda$CDM model (solid blue line with dotted blue lines showing 1$\sigma$ variances). The datapoints were rescaled by the $\beta$ parameter (see text). The horizontal solid black line is the Local Group velocity with respect to the CMB. Vertical lines give the limits of the 2MRS and 2MASS XSC catalogs. Note the general consistency of the observations with theoretical expectations.}
\label{Fig:g_m and v_c}
\end{figure}

\ind The most important result of this Paper is presented in Figure~\ref{Fig:g_m and v_c}. It shows the observed growth of the 2MASS dipole compared with the conditional velocity calculated from Eq.\ (\ref{eq:v_cond}). As was already discussed at the end of Subsec.\ \ref{Subsec:Theory}, we rescale observational data by a best-fit parameter $\beta\equiv\Omm^{0.55}\slash b$. The errorbars of the measured dipole were obtained from 1000-fold bootstrap resampling of the catalog. The $\beta$ parameter was fitted by a minimum-$\chi^2$ procedure, including both observational errors (from bootstrapping) and the theoretical (cosmic) variance. Note that formally we are `over-fitting', as the datapoints are correlated with each other, due to an integral nature of the measured dipole.

The observed growth, once rescaled, is well within $1\sigma$ range of the theoretical prediction (apart from the datapoints for the smallest distances, where the number of galaxies is very small, hence the measurement is noisy). Remarkably, the best agreement is obtained in the range $100\Mpch\lesssim r_\mathrm{eff}\lesssim300\Mpch$, i.e.\ between the extent of the 2MRS catalog and the 2MASS XSC completeness limit. The result of the fitting gives $\beta=0.38$ with $1\sigma$ confidence intervals of $\pm0.02$. Owing to the considerations of Sec.\ \ref{Sec:Data} concerning the scatter introduced by the offset added when passing from isophotal fiducial elliptical aperture magnitudes to total fluxes, as well as due to findings of \cite{BC10} on the bias introduced by the fact of possibly improper masking of the Local Void, we double this formal error. Hence our estimate is
\begin{equation}
\beta=0.38\pm0.04\;.
\end{equation}
This value is in agreement with the analysis of \cite{Erdogdu}, where also data from 2MASS were used, although from much smaller scales. The clustering dipole was calculated there from a subsample of the 2MASS XSC with measured redshifts, namely 2MASS Redshift Survey, then complete up to $\sim100\Mpch$ (maximum magnitude in the $K_s$ band equal to $11.25$ mag). Their result, based on claimed convergence of the flux-weighted dipole, was $\beta=0.40\pm0.09$. A somewhat larger value was obtained by \cite{PH}, where the peculiar velocity field within $65\Mpch$ predicted from 2MASS photometry and public redshift data was compared to three independent peculiar velocity surveys based on Type Ia supernovae, surface brightness fluctuations in elliptical galaxies, and Tully-Fisher distances to spiral galaxies. The best-fit from this comparison was $\beta=0.49\pm0.04$. On the other hand, our value of $\beta$ agrees within the errors with that obtained by \cite{Davis11}, who reconstructed the cosmological large scale flows in the nearby Universe using the SFI++ sample of Tully-Fisher measurements of galaxies and compared it with the whole sky distribution of galaxies in the 2MRS to derive $0.28<\beta<0.37$ (68.3\% confidence). Finally, in a recent paper, \cite{NBD11b} used the galaxy distribution in the 2MRS to solve for the peculiar velocity field, estimated absolute magnitudes of galaxies and constrained $\beta$ by minimizing their spread. They found a value in excellent agreement with our determination: $\beta=0.35\pm0.1$.

\section{Summary and conclusions}\label{Sec:SummConcl}
The motion of the Local Group of galaxies through the Universe can be used as a tool to constrain cosmological parameters. One of the applications is to estimate the peculiar acceleration of the LG and compare it within the linear theory with its peculiar velocity, known from the dipole anisotropy of the cosmic microwave background. This comparison may be used to directly measure $\beta=\Omm^{0.55}/b$, and hence the cosmological parameter of matter density, $\Omm$.

In this paper, as the estimator of the LG acceleration we have used the clustering dipole of the galaxies from the 2MASS Extended Source Catalog, which contains positions and fluxes of almost one million sources. As we wanted to reach as far from the Local Group as possible with the state-of-the-art observations, we have decided to trade the advantages of redshift surveys for the huge number of objects and unprecedented sky coverage of the 2MASS XSC. This allowed us to measure the dipole up to $300\Mpch$, which is almost 3 times farther than for the deepest existing all-sky survey of galaxies with measured redshifts, the 2MASS Redshift Survey (2MRS, a subset of the 2MASS XSC, containing $\sim$ 43,000 galaxies). The price to pay was our inability to weight galaxies with the inverse of their selection function (which would require redshifts to be known), so we could not directly mimic an `ideal', distance-limited survey, with our data. On the other hand, the flux dipole does not suffer from redshift-space distortions, and in particular from Kaiser's rocket effect.

We have focused mainly on the issue of convergence of the dipole, which has been a subject of debate for more than two decades now. We have shown that the flux dipole of the 2MASS XSC does \textit{not} converge up to the completeness limit of the sample, which is 13.5 mag in the $K_s$ band. Moreover, we find that beyond an effective distance of some $150\Mpch$, the growth is induced mainly in the negative Galactic Cartesian $y$ direction. We have checked that this behavior is not due to systematic effects related to masking out and artificially filling the Zone of Avoidance. Additionally, the misalignment angle between the measured dipole and the velocity of the LG is found to be of the order of $20\dgr$. We also find that the misalignment comes mainly from some nearby galaxies, bright in the near infrared. This is a hint that their removal could optimize the analysis, as was already theoretically investigated by \cite{CCBCC}. We plan to come back to this issue in the future.

Our inability to weight galaxies and to mimic a distance-limited survey and the lack of convergence of the dipole do not allow us to measure the parameter $\beta$ from direct $\bmv_{LG}$ -- $\bmg_{LG}$ comparison. However, by comparing the observed \emph{growth} of the clustering dipole with theoretical expectations for known $\bmv_\mathrm{CMB}$, in the framework of \cite{JVW} and \cite{LKH}, we find consistency with the predictions of the $\Lambda$CDM model, once the proper observational window of the 2MASS flux-limited catalog has been included. This fact enables us to measure the $\beta$ parameter from this comparison alone and the result is
\begin{equation}
\beta=0.38\pm0.04\;.
\end{equation}
This value is in agreement with and has a better precision than an earlier analysis, which equally used data from 2MASS, although from smaller scales: \cite{Erdogdu} found $\beta=0.40\pm0.09$ from the clustering dipole of the 2MASS Redshift Survey, with a depth of $\sim100\Mpch$.

Having estimated the value of $\beta$, we can go further and derive the density parameter $\Omm$  provided that we know the linear bias, $b$. The latter was estimated by \cite{Maller05}, who calculated the angular correlation function of galaxies in 2MASS and inverted it using singular value decomposition to measure the three-dimensional power spectrum. A fit of CDM-type power spectra in the linear regime allowed them to constrain the $K_s$-band linear bias as $b_K=1.1\pm0.2$. Using this value, we obtain the following estimate of the density parameter:
\begin{equation}
\Omm=0.20\pm0.08\;.
\end{equation}
This result, supporting the fact that the density of non-relativistic matter in the Universe is well bellow the critical one, is consistent with other independent dynamical estimates of $\Omm$, such as e.g.\ that by \cite{pairs}, who used mean relative peculiar velocity measurements for pairs of galaxies and obtained $\Omm=0.30^{+0.17}_{-0.07}$.

Our findings can be verified in at least two ways. The first would be to examine the growth of the 2MASS dipole with redshifts of galaxies as their distance estimates, using proper weights. In the near future we cannot however hope for a uniform and deep enough sample of spectroscopic redshifts for the catalog, even if the 2MASS Redshift Survey is continued (as it was planned to reach $K_s=12.25$ mag, see \citealt{Huchra}). The calculation may be thus feasible only for photo-\textit{z}'s, if they are available for the whole 2MASS XSC. A promising direction towards this goal may be to cross-correlate 2MASS with the data from the Wide-field Infrared Survey Explorer \citep[WISE,][]{WISE} that are currently partially available \citep{WISErelease} and are expected to be fully released in 2012. The other possible verification method could be for instance to examine the behavior of the differential dipole, i.e.\ to compare the increments of the growth with theoretical predictions of the $\Lambda$CDM model.

\section*{Acknowledgments}
The authors would like to thank Adi Nusser and Marc Davis for useful comments concerning an earlier version of this manuscript, as well as to the referee for valuable input. This publication makes use of data products from the Two Micron All Sky Survey, which is a joint project of the University of Massachusetts and the Infrared Processing and Analysis Center/California Institute of Technology, funded by the National Aeronautics and Space Administration and the National Science Foundation; the NASA/IPAC Extragalactic Database (NED), which is operated by the Jet Propulsion Laboratory, California Institute of Technology, under contract with the National Aeronautics and Space Administration and of the NASA/IPAC Infrared Science Archive, which is operated by the Jet Propulsion Laboratory, California Institute of Technology, under contract with the National Aeronautics and Space Administration. We acknowledge the use of the TOPCAT software, \texttt{http://www.starlink.ac.uk/topcat/} \citep{TOPCAT}. This research was partially supported by the Polish Ministry of Science and Higher Education under grants nos.\ N N203 0253 33 (MB \& MC) and N N203 509838 (MB). Part of this work was carried out within the framework of the European Associated Laboratory ``Astrophysics Poland-France'' (MB, MC \& GAM).

\appendix

\section{Effective distance for a given flux}\label{App:r_eff}
In this Appendix we show how we calculate the effective distance of galaxies with a given flux $S$, knowing their luminosity function (LF) in the given band, $\Phi(L)$, but not knowing their redshifts. Let us start by deriving the \emph{effective depth} of a flux-limited sample.

The number of galaxies $\de\mathcal{N}$ in a volume element $\de V=4 \,\pi\,r^2\, \de r$ of a spherical sample limited by minimum flux $S_\mathrm{min}$ (equivalent in the $K$ band to some limiting magnitude $K_\mathrm{max}$) is given by
\begin{equation}\label{eq:mathcalN}
\de\mathcal{N}=4 \,\pi\,r^2\, \de r\int_{4\pi r^2 S_\mathrm{min}}^{+\infty}\Phi(L)\de L\;,
\end{equation}
where $L_\mathrm{min}=4 \,\pi\,r^2\,S_\mathrm{min}$ is the minimum luminosity of galaxies in the sample at a given distance $r$. The \textit{mean depth} of the sample is defined as
\begin{equation}
\langle R \rangle=\frac{\int_0^{+\infty}{r\, \frac{\de\mathcal{N}}{\de r}}\,\de r}{\int_0^{+\infty}{ \frac{\de\mathcal{N}}{\de r}}\,\de r}\;,
\end{equation}
which, for the Schechter LF \citep{Sche} with a faint-end slope $\alpha$ and characteristic luminosity $L_*$, gives
\begin{equation}\label{eq:mean.depth}
\langle R \rangle=\frac{\int_0^{+\infty}r^3\, \Gamma(1+\alpha,4 \,\pi\,r^2\, S_\mathrm{min}\slash L_*)\,\de r}{\int_0^{+\infty}r^2\, \Gamma(1+\alpha,4 \,\pi\,r^2\, S_\mathrm{min}\slash L_*)\,\de r}\;,
\end{equation}
where $\Gamma(a,x)$ is the upper incomplete Gamma function.

Using the distribution (\ref{eq:mathcalN}), we can also easily derive the \emph{median depth} of a flux-limited sample, which we shall denote as $\bar{R}$. We obtain it by solving the integral equation:
\begin{equation}\label{eq:median.depth}
\int\limits_0^{\bar{R}}\frac{\de\mathcal{N}}{\de r}\,\de r=\int\limits_{\bar{R}}^{+\infty} \frac{\de\mathcal{N}}{\de r}\,\de r\;.
\end{equation}

The following Table \ref{Tab:depth} presents the mean and median depth of the 2MRS ($K_\mathrm{max}=11.25$~mag) and 2MASS XSC ($K_\mathrm{max}=13.5$~mag) for the LF as given by \cite{6dF_Fi}. The difference between the mean and the median is in that case of the order of a few percent.

\begin{table}[!h]
\begin{center}
\begin{tabular}{|l | c | c|}\hline
survey & mean depth $\langle R \rangle$ & median depth $\bar{R}$ \\ \hline
2MRS & $82\Mpch$  & $76\Mpch$ \\ \hline
2MASS XSC & $232\Mpch$ & $215\Mpch$ \\ \hline
\end{tabular}
\end{center}
\caption{\label{Tab:depth}Mean and median effective depth of flux-limited surveys: the 2MASS Redshift Survey (2MRS, $K_\mathrm{max}=11.25$ mag) and the 2MASS Extended Source Catalog (XSC, $K_\mathrm{max}=13.5$ mag), calculated from Eqs.\ (\ref{eq:mean.depth}) and (\ref{eq:median.depth}), respectively, for the $K$-band luminosity function in the Schechter form with $\alpha=-1.16$ and $M_*=-23.83+5\log h$, \cite{6dF_Fi}.}
\end{table}

The effective depth of a flux-limited survey is however not a good measure of the effective \textit{distance} of a thin shell in `flux space' (in which the third coordinate is the flux, by analogy with the redshift space) that we seek. What we need is the mean value of distances of all galaxies \textit{with given flux} $S$. The mean that we calculate will be thus a conditional one. This derivation is qualitatively the same as for the mean redshift  of galaxies with a given flux, presented on pages 120--121 of \cite{Pe93}.

We start by deriving the joint probability distribution of galaxy distances $r$ and fluxes $S$. It is easily obtained by differentiating Eq.\ (\ref{eq:mathcalN}) with respect to luminosity $L$. Hence, the differential number $\delta^2 N$  of galaxies with a LF $\Phi(L)$ in a volume element $\delta V$ is given by
\begin{equation}
\delta^2 N=\Phi(L)\,\delta L\,\delta V
\end{equation}
with $\delta V=4\pi r^2\,\delta r$. Now, passing from luminosity to flux, $L=4\pi r^2 S$, for fixed $r$ we have $\delta L=4\pi r^2\, \delta S$. This gives the joint probability as
\begin{equation}
p(r,S)=\frac{\partial^2 N}{\partial r\, \partial S}=16\,\pi^2\,r^4\,\Phi(4\pi r^2 S).
\end{equation}
The conditional probability for $r$ given $S$ is
\begin{equation}
p(r|S)=\frac{p(r,S)}{p(S)}=\frac{16\,\pi^2\,r^4\,\Phi(4\pi r^2 S)}{\int_0^{+\infty}{16\,\pi^2\,r^4\,\Phi(4\pi r^2 S)}\,\de r}\;,
\end{equation}
hence the \emph{conditional mean} for $r$ given $S$ will be
\begin{equation}\label{eq:r_mean}
\langle r \rangle_S=\frac{\int_0^{+\infty}{r^5\,\Phi(4\pi r^2 S)\,\de r}}{\int_0^{+\infty}{r^4\,\Phi(4\pi r^2 S)}\,\de r}\;.
\end{equation}
For the Schechter form of the luminosity function, let us define a characteristic distance $r_S$, which is a distance to a galaxy with a given flux $S$ and a characteristic luminosity $L_*$:
\begin{equation}
r_S=\sqrt{\frac{L_*}{4\pi S}}\;.
\end{equation}
Then for the Schechter LF, Eq.\ (\ref{eq:r_mean}) simplifies to
\begin{equation}
\langle r \rangle_S=
r_S\,\frac{\Gamma(\alpha+3)}{\Gamma(\alpha+5\slash 2)}
\end{equation}
with $\Gamma(a)$ being the Gamma function.

Similarly, the \emph{conditional median} for $r$ given $S$, denoted here as $\bar{r}_S$, will be obtained from the implicit equation
\begin{equation}
{\int\limits_0^{\bar{r}_S}{r^4\,\Phi(4\pi r^2 S)}\,\de r}={\int\limits_{\bar{r}_S}^{+\infty}{r^4\,\Phi(4\pi r^2 S)}\,\de r}\;,
\end{equation}
which for the Schechter LF is equivalent to solving
\begin{equation}
\gamma\left(\alpha+\frac{5}{2},\frac{\bar{r}_S^2}{r_S^2}\right)=\Gamma\left(\alpha+\frac{5}{2},\frac{\bar{r}_S^2}{r_S^2}\right)
\end{equation}
with respect to $\bar{r}_S$, where $\gamma(a,x)$ and $\Gamma(a,x)$ are respectively the lower and upper incomplete Gamma functions.

In the particular case of the $K$ band, taking $\alpha=-1.16$ \citep{6dF_Fi}, we obtain
\begin{equation}
\langle r \rangle_S=1.056\, r_S
\end{equation}
and
\begin{equation}
\bar{r}_S=1.013\, r_S\;.
\end{equation}
Note that especially the median conditional distance gives a value very close to the characteristic distance $r_S$, which could be the `first-guess' effective distance (Tully 2008, private communication).

For the purpose of the present paper, we have decided to use the conditional median as our measure of the effective distance of galaxies with a given flux. This particular choice is somewhat arbitrary and influences scalings of top axes in Figs.\ \ref{Fig:growth.r}, \ref{Fig:misal}, \ref{Fig:predicted} \& \ref{Fig:g_m and v_c}. However, since the difference between the conditional mean and median is very small (about 5\%), this choice has a negligible impact on general conclusions and results of this paper.

Relating the flux to the $K$ magnitude via Eq.\ (\ref{eq:S_i}), $S=S_0\,10^{-0.4 K}$, and owing to the luminosity-magnitude relation $L_*=4 \pi (10\, \mathrm{pc})^2 S_0\,10^{-0.4 M_*}$, we obtain
\begin{equation}
r_\mathrm{eff}\equiv \bar{r}_S=1.013\times 10^{0.2(K-M_*)-5}\,\mathrm{Mpc}\;,
\end{equation}
so finally, for $M_*=-23.83+5\log h$ \citep{6dF_Fi},
\begin{equation}\label{eq:r.eff.app}
r_\mathrm{eff}=0.591\times10^{0.2K}\Mpch\;.
\end{equation}
For example, galaxies at the limit of the 2MASS catalog (with $K=13.5$ mag) are assigned an effective distance of $r_\mathrm{eff}=296\Mpch$.
\bibliographystyle{apj}
\bibliography{growth}
\end{document}